\documentstyle[11pt,aaspp4,flushrt]{article}

\slugcomment{
}
\lefthead{K. M. Xilouris et al.}
\righthead{The characteristics of millisecond pulsar emission II.}

\begin{document}

\title{The characteristics of millisecond pulsar emission: \\
II. Polarimetry}

\author{Kiriaki M. Xilouris\altaffilmark{1}, 
Michael Kramer\altaffilmark{2},
Axel Jessner\altaffilmark{2},
Alexis von Hoensbroech\altaffilmark{2},
Duncan Lorimer\altaffilmark{2},
Richard Wielebinski\altaffilmark{2},
Alexander Wolszczan\altaffilmark{3},
Fernando Camilo \altaffilmark{4}}

\altaffiltext{1}{National Astronomy and Ionosphere Center, Arecibo Observatory, P.O. Box 995, Arecibo, PR 00613, USA}
\altaffiltext{2}{Max-Planck-Insitut f\"ur Radioastronomie, Auf dem H\"ugel 69,
53121  Bonn, Germany} 
\altaffiltext{3}{Department of Astronomy and Astrophysics, 
Penn State University, University Park, PA 16802, USA} 
\altaffiltext{4}{Nuffield Radio Astronomy Laboratories,
Jodrell Bank, Macclesfield, Cheshire SK11 9DL, England; Marie Curie Fellow
}

\begin{abstract}

We have made polarimetric monitoring observations of most of the
millisecond pulsars visible from the northern hemisphere at 1410 MHz
over a period of three years.  Their emission properties are presented
here and compared with those of normal pulsars. 
Although we demonstrated in paper I that millisecond pulsars
exhibit the same flux density spectra and similar profile complexity, 
our results presented here suggest
that millisecond pulsar profiles do not comply with the
predictions of classification schemes based on ``normal'' pulsars.
The frequency development of a large number of millisecond pulsar
profiles is {\em abnormal} when compared with the development seen for
normal pulsars.  Moreover, the polarization characteristics suggest
that millisecond-pulsar magnetospheres might not simply represent
scaled versions of the magnetospheres of normal pulsars,
supporting results of paper I.  However,
phenomena such as mode-changing activity in both intensity and
polarization are recognized here for the first time (e.g.,
J1730--2304).  This suggests that while the basic emission mechanism
remains insensitive to rotational period, the conditions that,
according to the canonical pulsar model, regulate the radio emission,
might be satisfied at different regions in millisecond pulsar
magnetospheres.

At least three types of model have been proposed to describe the
millisecond pulsar magnetospheres, ranging from distorted magnetic
field configurations due to the recycled nature of these sources to
traditional polar-cap emission and emission from outer gaps.  A
comparison of the predictions of these models with the observations
suggests that individual cases are better explained by different
processes. However, we show that millisecond pulsars can be grouped
according to common emission properties, a grouping that 
awaits verification from future multifrequency observations.

\end{abstract}
\section{Introduction}

We have been monitoring the polarization characteristics of 24
millisecond pulsars (MSPs) at 1410 MHz for a period of three years,
aiming at contrasting their emission properties and those of normal
pulsars.  Previous studies of individual MSPs have shown that their
emission properties are complex (e.g., \cite{tho90}; \cite{man92};
\cite{man95}; \cite{nav96}; \cite{arz96}).  This work
provides the first approach in determining the systematics involved in
the MSP emission properties, and compares them with the predictions of
schemes based upon studies of normal pulsars (e.g., \cite{R83a} and
\cite{lyn88b}).

At least three types of theoretical models have been developed to
describe the magnetic field topology in MSP magnetospheres. This is a
topic of considerable interest due to its impact on the evolutionary
history and the statistics of this stellar population.  Some models
assume unusual magnetic field configurations for MSPs, due to the
recycling which occurs during their evolution (\cite{rud91};
\cite{che93b}).  Other models constrain the non-dipolar magnetic field
topology, by postulating vacuum electromagnetic radiation
(\cite{kro91}). In addition, the presence of higher order multipoles is
severely constrained by the location of the spin-up lines as
derived from magnetic torques (\cite{aro93}).  A third class of model
postulates emission resulting from outer magnetospheric gaps (e.g.,
Cheng, Ho \& Ruderman 1986; \cite{che93a}; \cite{rom95}), or even a
contribution of emission from traditional polar caps in addition to
this process.  While taking different approaches, these models
sometimes predict only slight differences in the emission properties
of the MSP magnetospheres.  The predictions of these models are traced
here using the largest sample of MSP polarimetry available so far.

The magnetic field configuration places constraints on pulsar
evolutionary models, a subject of particular interest following the
discovery of MSPs.  High-quality polarimetric data potentially provide
a means for determining the observed viewing geometry of a pulsar,
namely the magnetic inclination $\alpha$ and the line-of-sight
trajectory $\beta$.  Observationally, it is uncertain whether the
magnetic inclination for normal pulsars evolves secularly on
time-scales of $\simeq10^{7}$ yr (e.g., \cite{lyn88b}; \cite{ran92};
\cite{bat92}; McKinnon 1993).  It is therefore a curious observation
that there is a high percentage of pulsars with interpulses (IPs) or
emission at low levels extending for a significant fraction of the
pulsar period (signatures of orthogonal and aligned orientation
respectively), among the MSP population, composed primarily of very
old objects.  However, to determine whether this observation is
significant, the systematics of MSP emission must first be well
understood.

As an essential step towards meaningfully interpreting MSP data, it is
vital to understand the systematics involved in their emission
properties (spectra, intensity fluctuations, polarimetry and profile
morphology).  The present work, accompanied by a study of the profile
shapes (\cite{paperI}; hereafter paper I) and the location of the
emission in MSP magnetospheres
(Xilouris {\it et al.} ~1998a; paper III),
provide the first such systematic approach.  In the following
sections, the emission properties of 26 MSPs in total,
24 in our sample and two based on the available literature,
are examined in the
context of polar-cap emission as it is understood from extensive
studies of normal pulsars (e.g.~\cite{ran92}; \cite{lyn88b}; Gould
1994).  The wide applicability of these classification schemes is
discussed in view of the complex morphology and emission properties
seen for MSP profiles (Section 3).  The new trends identified in MSP
profiles are summarized in Section 4, followed by a statistical
analysis contrasting the properties of MSP and normal pulsars.  The
impact of physical conditions on MSP emission is examined in Section
5, where our results are discussed in the context of current
theoretical models that attempt to describe MSP magnetospheres.

\section{Millisecond-Pulsar Monitoring}

Millisecond pulsar monitoring at 1410 MHz was initiated at the
Effelsberg radio-telescope in April 1994 
following a
major upgrade of the associated data-acquisition hardware. These
modifications were made to permit high time resolution observations
(down to 0.2$\mu$s), and to increase the system stability, as required
to study short-period pulsars.  An adding polarimeter was used which
accepts opposite-hand circularly polarized inputs at an intermediate
frequency of 150 MHz, with a bandwidth of 40 MHz, through a network of
hybrids followed by an on-line de-disperser.  A full description of
the polarimeter, the gain stability and the polarimetric calibration
is presented by von \cite{hoe97}.

During the three-year monitoring period, most of the sources presented
in this work were routinely observed in 31 sessions, spaced about a
month apart. During each run, the polarization properties of some
well-studied pulsars (PSRs B1929+10, B0540+23, B1937+21 and B1855+09)
were observed to enable the monitoring of instrumental cross-coupling.
By incrementing the position angle of the $\lambda$21---18 cm feed at
Effelsberg, an offset parallactic angle is added to the input
signal. A rapid variation of the {\em effective parallactic angle}
(true parallactic angle + offset) is obtained in this way. This
feature permits a quick determination of the instrumental
cross-coupling phase and amplitude. Gain monitoring was achieved by
injecting a calibration signal asynchronously with the pulsar period
both before and after each scan.  An extensive discussion of the
calibration procedure is presented elsewhere (von ~Hoensbroech \&
~Xilouris 1997).

\begin{figure}
\plotfiddle{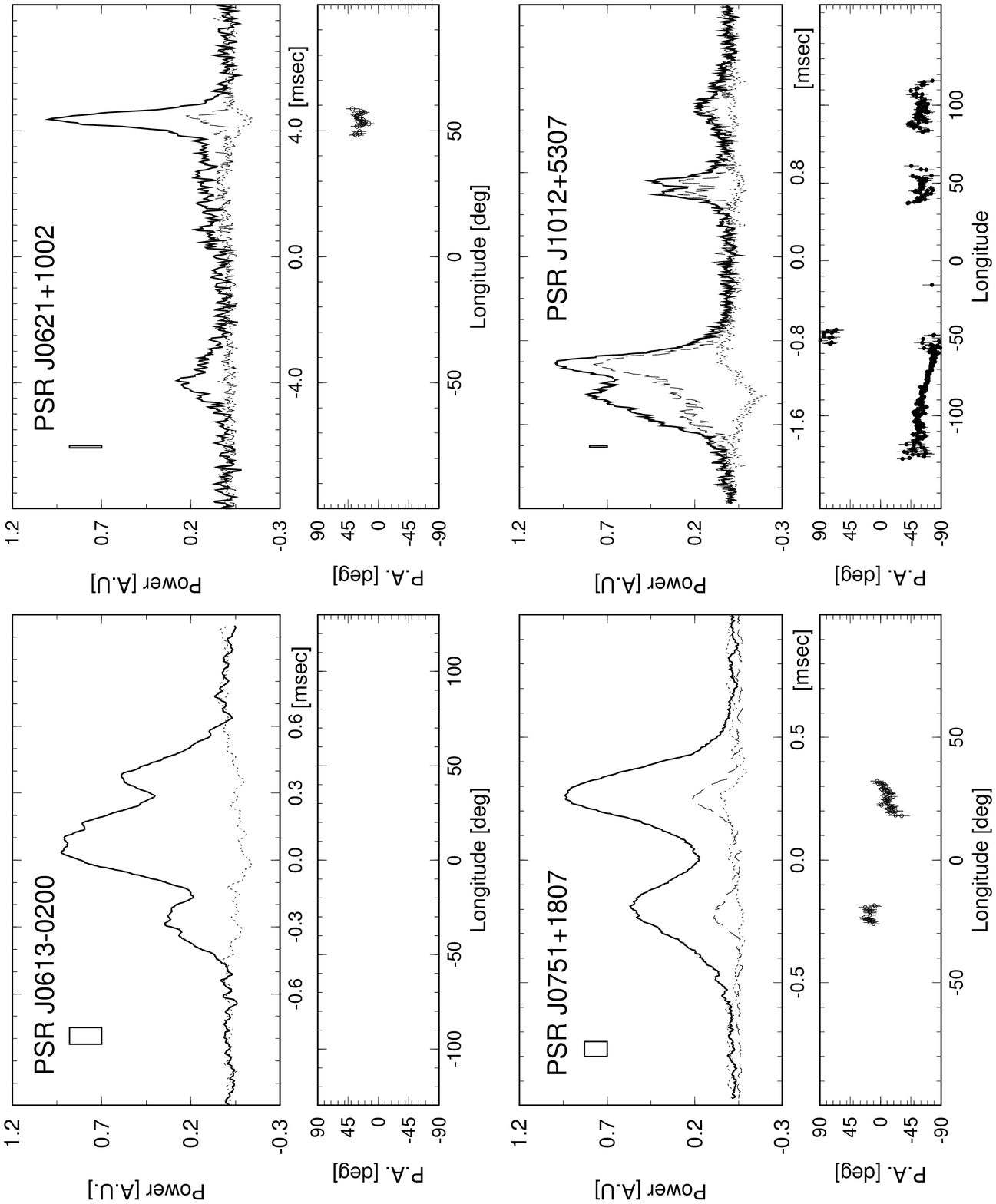}{10cm}{270}{60}{60}{-250}{350}
\bigskip

\figcaption{ \label{plt_1} 1410 MHz polarization profiles of 23 MSPs.
Four curves are plotted: the total power (Stokes parameter I)
is the outer curve, whose maximum is scaled to arbitrary units. The
linear polarization (Stokes $L=(Q^2+U^2)^{1/2}$) is the interior,
dashed curve, and the circular polarization (Stokes $V=$LCP$-$RCP) is
the dotted curve. The linear PPA curve is plotted on the lower panel
in degrees and measured counter-clockwise from north on the plane
of the sky. 
The PPA curve is plotted only when the linear polarization is greater
than 2$\sigma$, except in the cases of PSRs J0613-0200, B1534+12 and
J2322+2057 where the threshold is 1.5 $\sigma$.
All four curves are plotted against the pulsar longitude in ms (upper
scale) and degrees (lower scale).
}
\end{figure}

\begin{figure}
\plotfiddle{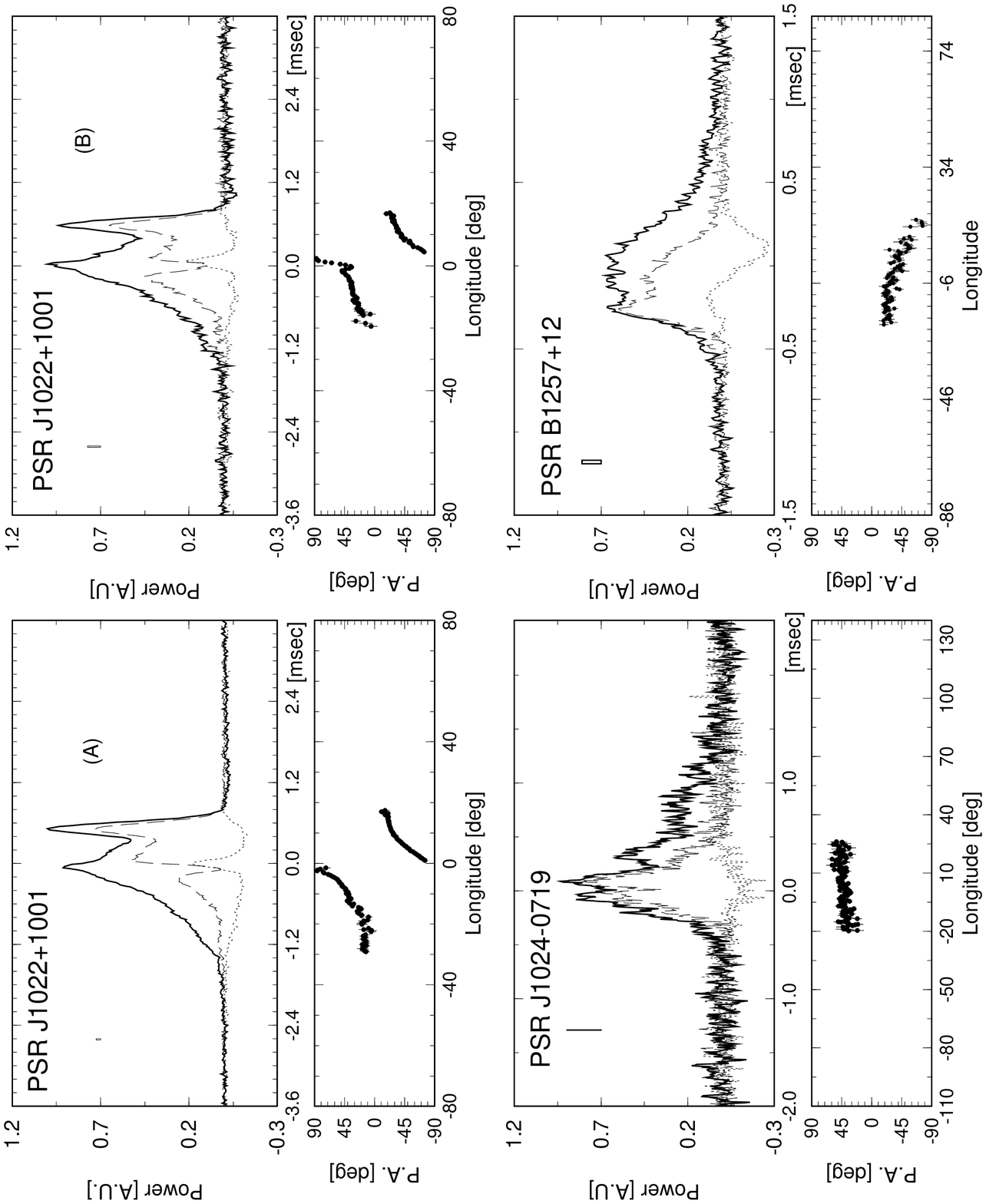}{10cm}{270}{60}{60}{-250}{350}
\bigskip

\figcaption{ \label{plt_2}
see Fig.~1
}
\end{figure}

\begin{figure}
\plotfiddle{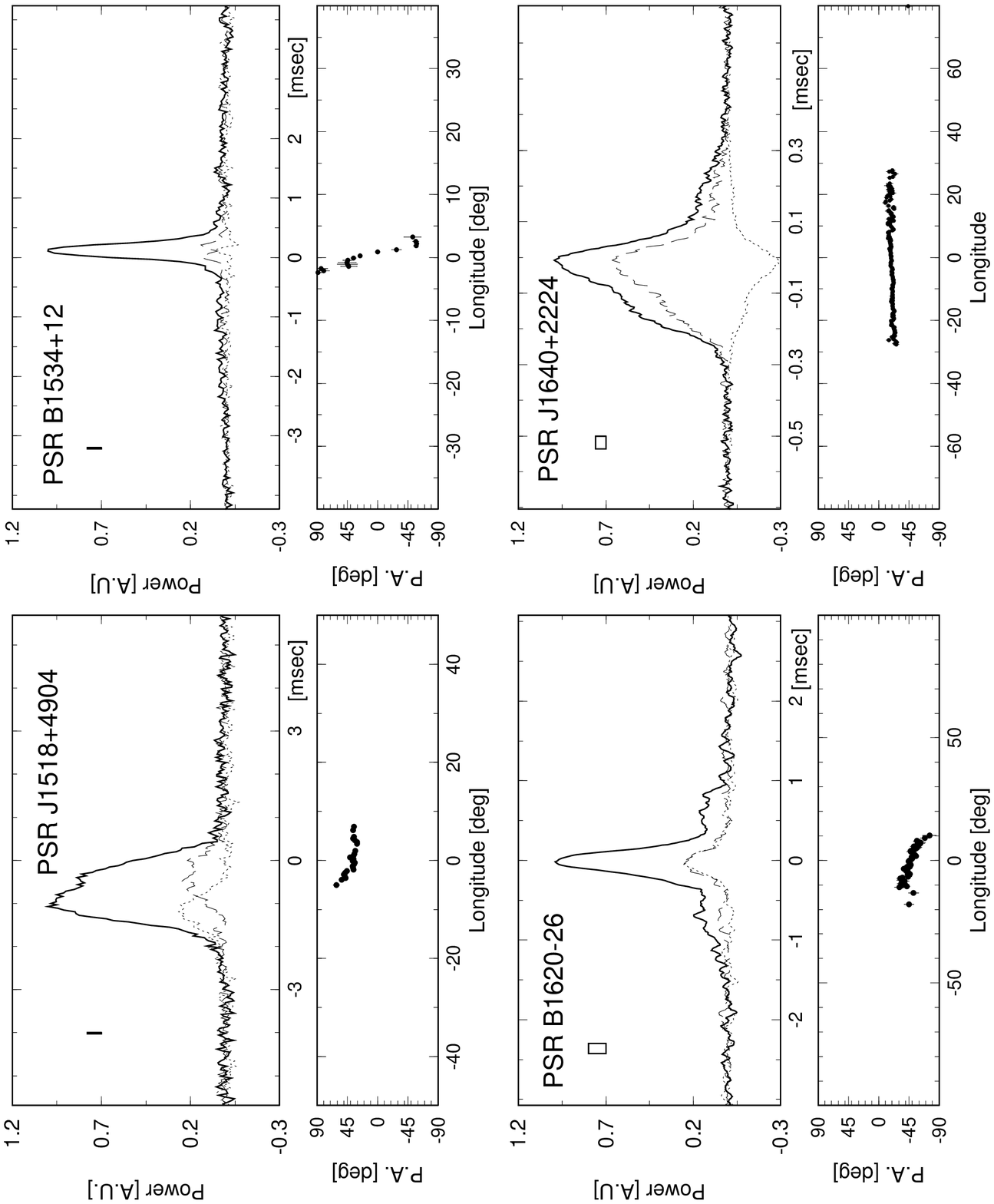}{10cm}{270}{60}{60}{-250}{350}
\bigskip

\figcaption{ \label{plt_3}
see Fig.~1
} 
\end{figure}

\begin{figure}
\plotfiddle{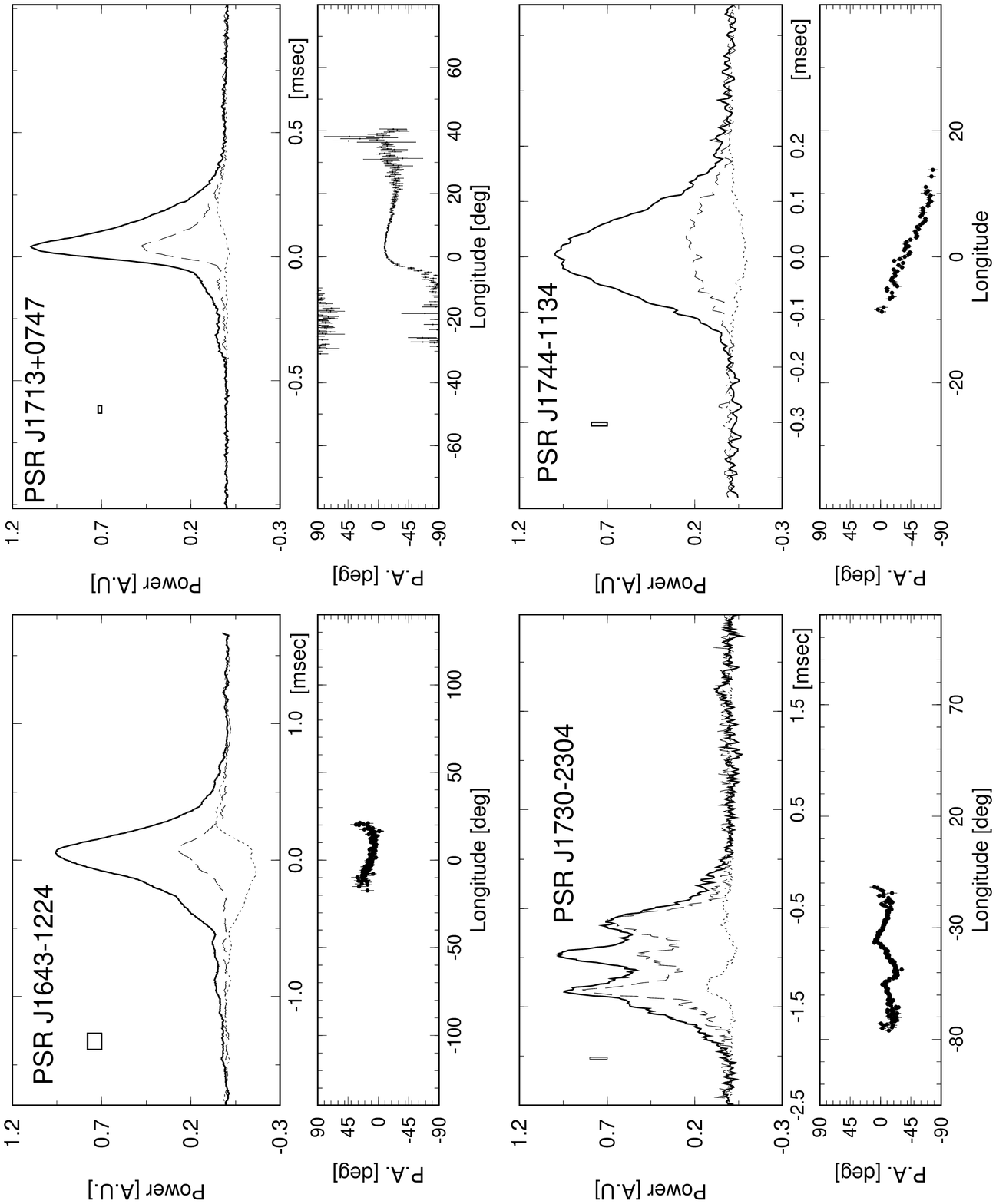}{10cm}{270}{60}{60}{-250}{350}
\bigskip

\figcaption{ \label{plt_4}
see Fig.~1
}
\end{figure}

\begin{figure}
\plotfiddle{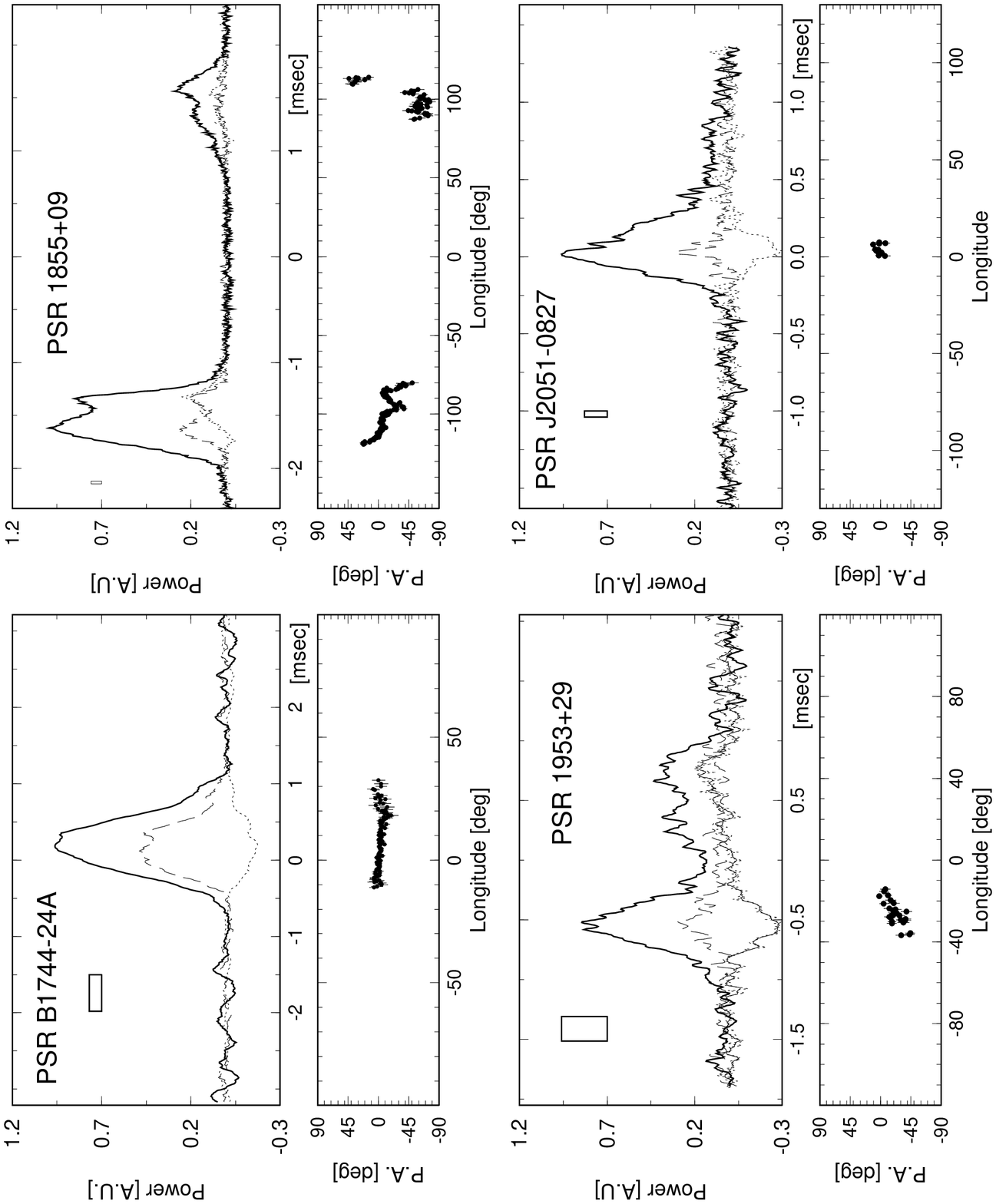}{10cm}{270}{60}{60}{-250}{350}
\bigskip

\figcaption{ \label{plt_5}
see Fig.~1
}
\end{figure}

\begin{figure}
\plotfiddle{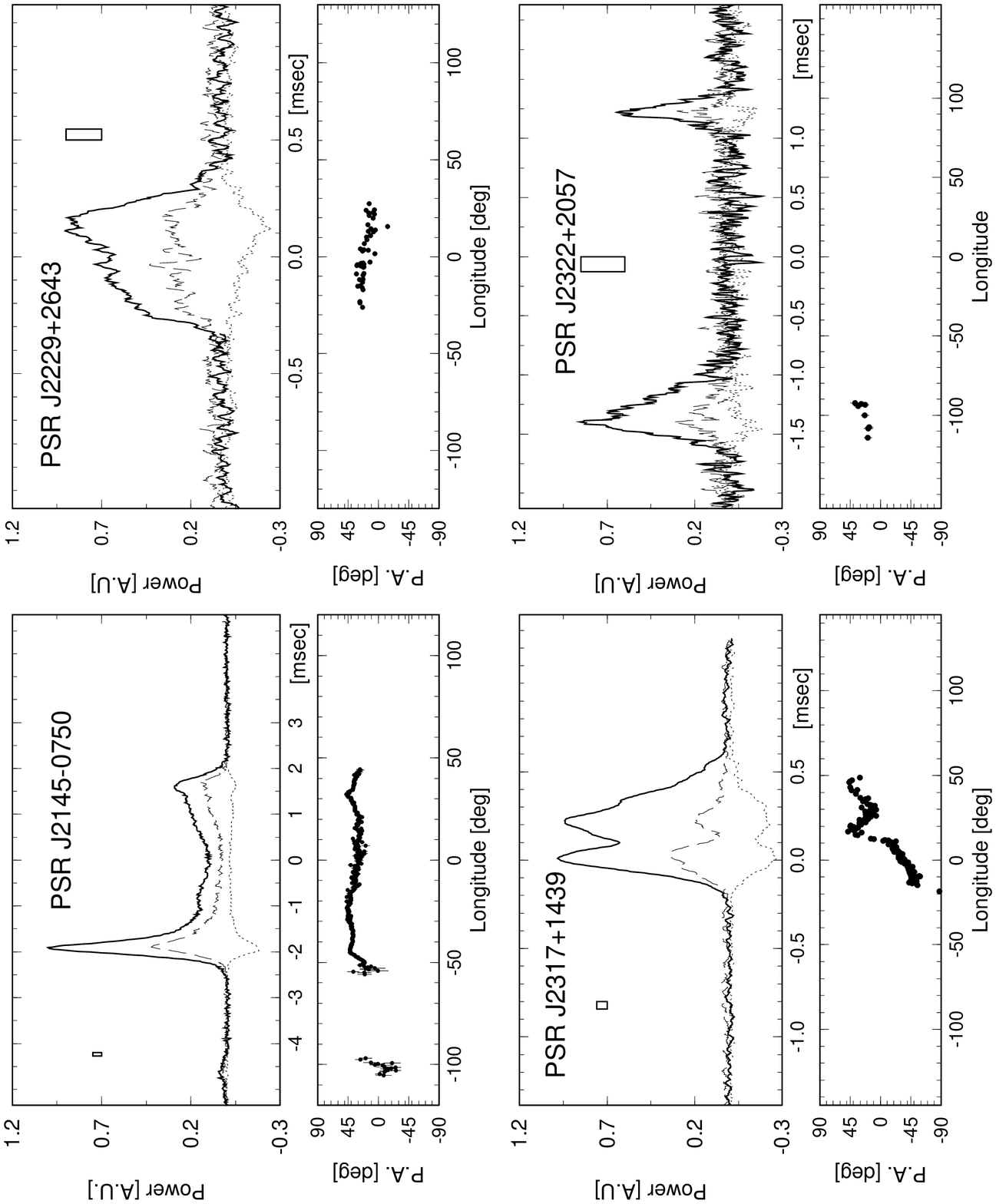}{10cm}{270}{60}{60}{-250}{350}
\bigskip

\figcaption{ \label{plt_6}
see Fig.~1
}
\end{figure}

The polarization profiles of most of the pulsars in our sample are
presented in Figs.~\ref{plt_1} through \ref{plt_6}.  Pulsars observed
in total power mode only, such as PSR J2019+2425, or badly smeared
profiles (e.g., B1937+21) are not presented here.  The dispersion
smearing is indicated by the resolution box on each profile. This time
delay is computed for the central channel (i.e., bandwidth of 0.666
MHz centred at 1410 MHz) of the 60 de-disperser channels.  The
smearing correction applied to each individual channel is computed by
a linear extrapolation of this central channel value.
 
\section{A study of individual pulsars}
 
As a pulsar rotates, different parts of its magnetosphere 
are swept across the observer's line of sight.
These parts are recognized as the components of emission that 
constitute the integrated profiles, which are known to exhibit 
considerable diversity in their morphology.
Profile diversity has been addressed by a simplified picture 
(Rankin 1983a, b) which has served to explain the
profile morphology of normal pulsars. Rankin proposes that two generic
component types exist which combine in a systematic fashion to explain
the diversity seen in normal pulsar profiles.  Lyne \& Manchester
(1988) confirm this trend, but postulate a random rather than a
deterministic process in which the components combine to form an
integrated profile. Both models recognize that the spectral
development as well as the polarization properties of the profile
components vary with the location of a component within the integrated
profile.  It has been shown by Rankin (1983a) and confirmed by Lyne \&
Manchester (1988) that there is generally a difference in spectral
index between the central and outer parts of the emission beam. The
central part, associated with the ``core''
component in Rankin's model, usually exhibits a steeper spectrum than
the outer parts (or ``conal'' components), and is
emitted from regions closer to the stellar surface (\cite{R83a}).

The level of circular polarization associated with the core component
is often higher than at other longitudes, and exhibits a sense
reversal within the core component.  This property greatly assists in
identifying the core components, features which are associated with
central trajectories of the observer's line-of-sight.  The
polarization position angle (PPA) of the linearly polarized component
of radio emission ideally follows a curve described by the rotating
vector model (Radhakrishnan \& Cooke 1969, hereafter the RVM).  In
practice, the PPA curves often undergo sudden discontinuities from
this predicted behavior (orthogonal modes) at certain pulse longitudes
which correlate with drops in the intensity of the linear
polarization.  In most cases, when these orthogonal modes are
accounted for (e.g.~Stinebring {\it et al}.~1984) a PPA curve can be
reconstructed and fitted with a curve described by the RVM.
Complicated field topologies due to the recycled nature of MSPs
(Ruderman 1991) and gravitational bending (e.g Pechenick {\it et al.}
1983; Chen \& Ruderman 1993b) are just some of the effects expected to
have influenced the MSP emission in near-surface emission models.
Similar effects are anticipated if the emission originates from the
proximity of the light cylinder. Here, magnetic sweep-back would
influence both the profile structure and the PPA curve (Barnard 1986).
In addition, the relativistic motion in a binary system is expected to
affect, in a systematic fashion, the profile shapes and polarization
characteristics (Damour \& Taylor 1992) of pulsars in such systems.

In this section, the polarization properties and profile development
summarized above are traced for each MSP in our sample. PSRs B1937+21
and B1957+20 are also discussed due to their special emission
properties, based on the literature available.  A quantitative
analysis of the polarization results is discussed in Section 4 and
summarized in Table \ref{tbl-1}.

\placefigure{plt_1}
\placefigure{plt_2}
\placefigure{plt_3}
\placefigure{plt_4}
\placefigure{plt_5}
\placefigure{plt_6}

\subsection{PSR J0613$-$0200}

This MSP (\cite{lor95}) is a member of a low-mass binary system in a
1.2 day circular orbit. Its companion is likely to be a white dwarf of
mass $\approx 0.15 M_{\sun}$.  The weak but systematic sense-reversing
circular polarization at the profile centre suggests that this is a
core component.  The presence of such a component implies that the
line of sight crosses the polar cap region near the magnetic axis,
making a small impact angle $\beta$.  At $\lambda$21 cm, normal
pulsars with triple profiles, particularly those with wide profiles,
preserve some amount of linear polarization, while PSR J0613--0200 is
essentially unpolarized. Very little frequency development is seen in
the profile morphology of this pulsar in contrast with the predictions
for a triple profile.  A basic triple symmetric structure is evident
over a wide range of frequencies (\cite{bel96}).  One exception is
evident in a low signal-to-noise profile at 780 MHz (\cite{bac95}),
where the trailing component appears abnormally bright in comparison
to the leading and core components.  This {\em abnormal} development
is not seen at higher frequencies.  Typically, for normal pulsars, the
central part of the profile exhibits a steeper spectrum than the outer
components, and is thus expected to fade faster towards higher
frequencies (\cite{R93a}).  The observed minimal spectral evolution of
the components is in contrast to the predictions of the Rankin model.

\subsection{PSR J0621+1002}

Being a member of a wide orbit binary system with a $\approx 0.5
M_{\sun}$ companion (Camilo {\it et al.}~1996a)
this pulsar belongs to the proposed class of intermediate-mass binary
pulsars. Its
profile is among the widest known, and has a relatively sharp
component towards the trailing edge, showing significant circular, and
moderate linear polarization suggesting that this component could
have been the profile core.  Due to its displaced position in the
profile, the classification of this component is therefore uncertain.
The PPA curve is shallow and does not follow a RVM curve.  The triple
symmetric profile morphology remains unchanged between 370 MHz
(Camilo {\it et al.}~1996a) and 1.4 GHz (Fig.~\ref{plt_1}),
indicating {\em slow} spectral evolution of the components.  The
narrowing of the profile over this frequency range is also negligible,
perhaps less than one degree.

\subsection{PSR J0751+1807}

This pulsar, recently detected at X--ray wavelengths (\cite{bec96}) is
in a short orbital period binary system, with a white-dwarf companion
of $0.15 M_{\sun}$ (\cite{lun95}).  At 1.4 GHz, J0751+1807 displays a
symmetric double-peaked profile.  While the profile shape was stable
throughout the three years of our observations, the polarization of
its leading component varied.  The linearly polarized intensity of the
leading component fluctuates, while the associated PPA curve changes
slope.  In contrast, the PPA associated with the trailing profile
component remains stable with time.  The amount of circular
polarization is small, and changes sense towards the trailing
component.  The profile of PSR J0751+1807 exhibits an {\em unexpected}
frequency development. The 430-MHz profile (\cite{lun95}) exhibits one
bright leading component.  In contrast to all expectations, at 1.4 GHz
the trailing part develops into a distinct component which is brighter
than the leading one.  The profile width at 430 MHz is
67$^{\circ}\pm$3$^{\circ}$, while we measure a width of
73$^{\circ}\pm$6$^{\circ}$ at 1.4 GHz. This is in good agreement with
the 1.4-GHz profile presented by ~Lundgren {\it et al.}~(1995), and
also suggests insignificant profile narrowing within the errors of
measurements.  At both 430 MHz and 1.4 GHz, low-level emission is
detected up to $\approx 130^{\circ}$ in pulse longitude.

\subsection{PSR J1012+5307}

This field MSP is a member of a binary system (\cite{nic95b}), in a
tight circular orbit with a low-mass ($0.15 M_{\sun}$) white-dwarf
companion (\cite{lor95b}; van \cite{ker96}).  It exhibits the most
perplexing profile yet seen with at least nine components making up
two or three separate structures (cf.~paper I).  A main pulse (MP) and
a first interpulse (IPa) that resembles the MP in shape and
polarization are followed by a third structure (IPb) which appears to
be superposed on IPa.  There is a strong resemblance between this
profile and that of PSR B1055$-$52, a normal pulsar that is almost
certainly an orthogonal rotator (\cite{R83a}).  However, such a
conclusion is far from obvious here, due to emission throughout most
of the pulsar period.  Moderate amounts of sense-reversing circular
polarization are associated with both the MP and IPa, indicating the
existence of core components in both structures. This suggests that
even though their separation is only $\Delta \phi_{\rm MP-IP} =
70^\circ$, they might result from a line of sight that crosses near
two magnetic poles which are located less than 180$^{\circ}$ apart.
Most likely, the MP and the IPb fit the classical MP -- IP
description, since their midpoint separation is close to
180$^{\circ}$. However, emission at low levels continues throughout
most of the pulsar period (e.g \cite{nic95b}), suggesting that all
structures could as well originate from a single pole.

Assuming that the opening semiangles $\rho$ of the beams of a MP and
an IP are the same, the impact parameter of the IP follows from
$\beta_{\rm IP} = 180 - 2\alpha - \beta_{\rm MP}$. Therefore, in the
case of an orthogonal rotator, one expects $\beta_{\rm
IP}=-\beta_{\rm MP}$.  This directly reflects on the maximum rate of
change of the PPA $(\frac {\sin \alpha} {\sin \beta})$.  Equal slopes
for the PPA, but of opposite sign, should thus indicate an orthogonal
rotator.  While this should be a simple test, helpful in determining
the viewing geometry of pulsars with IPs, its value is questionable
due to the ambiguity seen in the slopes of normal pulsars with IPs.
In this case, equal slopes of the same sign for the MP and IPa, would
indicate an aligned rotator, although we cannot exclude an ambiguous
classification.

Observations of lower resolution at three frequencies (\cite{nic95b})
demonstrate an {\em abnormal} frequency development, with IPa
exhibiting a much shallower frequency dependence than the outer
structures.  There is also little frequency dependence in the
component separations, which is a property that has been previously
regarded as indicating emission from different magnetic poles
(\cite{han86}).  The low mass of the spectroscopically detected
white-dwarf companion suggests an orbital inclination of
$\approx 67 ^\circ$, close to an edge-on geometry.  Assuming the
angular-momentum axes of the binary system and the pulsar rotation are
aligned, the suggested geometry is consistent with an orthogonal
rotator. Such a geometry is further supported by RVM curve fits to the
PPA of this pulsar, where $\alpha  \approx 88 ^\circ$, and
$\beta  \approx 5 ^\circ$ have been derived (paper III).
 
\subsection{PSR J1022+1001}

This pulsar is in a wide circular orbit with a $\approx 0.87 M_{\sun}$
companion (\cite{cam96b}).  The emission is confined to a relatively
narrow symmetric profile within which four or five sharp components
can be fitted (paper I).  Substantial circular polarization that
exhibits a sense reversal identifies the central part of the profile
with a core component.  The linear polarization varies across the
profile from essentially zero for the leading part, to a highly
polarized, sharp trailing component.  Profile and polarization
variations have been identified where the leading and the trailing
component alternate in strength.
The relative intensity of the two components is seen to evolve
at a rate inconsistent with pulsar mode changing, leading to
two distinct profiles.
The most common profile where the trailing component
appears stronger than the leading component is shown in Fig. 2
annotated as (A), while the less frequent one which also
exhibits a much sharper PPA is annotated as (B).
The PPA curve is often disturbed by
orthogonal mode-changing.  However, when these discontinuities are
accounted for by applying appropriate shifts, the PPA closely follows
an RVM curve. A fit of the RVM curve to the PPA curve for this pulsar
yields $\alpha \approx 53.2 ^{\circ}$, and $\beta \approx 7.3
^{\circ}$ (paper III).  From this viewing geometry, we were able to
estimate the emission height of the radio-emitting region. Assuming a
dipolar magnetosphere, the height is consistent with 3.2
$R_{\ast}$. At 430 MHz, pulse shape variations are also seen, while at
1.7 GHz, the leading component is brighter most of the time (~Camilo
et al. 1998). At 4.85 GHz (~Kijak {\it et al.}~1997), the trailing
component has faded with respect to the leading one.  Between 430 MHz
and 4.85 GHz, the profile development is {\em not consistent} with
that of a triple or multiple profile.  It is atypical of MSP profiles
to exhibit the secular variations found in this source.

\subsection{PSR J1024$-$0719}

This recently discovered isolated pulsar (\cite{bai96}) exhibits a
multi-component profile with high linear and moderate amounts of
circular polarization associated with the central component.
Comparing with lower frequency profiles (\cite{bai96}), the central
component has gained in intensity at higher frequencies.  If this is
associated with the core component, then the spectral evolution is
{\em contrary} to that expected for a pulsar of this type.
 
\subsection{PSR B1257+12}

This pulsar, the central object of the first extra-solar planetary
system to be discovered (\cite{wol92}), exhibits a very high degree of
circularly polarized emission with sense reversal. This indicates the
presence of a core component in the centre of the profile, implying
that the line-of-sight cuts the polar cap close to the magnetic
axis. The high degree of linear polarization is consistent with the
highly polarized core components seen in the core-single profiles of
normal pulsars.  There is no significant narrowing of the profile
between 430 MHz (\cite{wol92}), 790 MHz (\cite{bac95}), and 1.4 GHz,
{\em nor is there a significant} morphological development. The core
component is brighter at 430 MHz, but has faded and becomes comparable
in intensity to the trailing profile component at 1.4 GHz.  This is in
good agreement with the development of a triple profile.  The PPA swing is
shallow but well defined, and in agreement with a dipolar
magnetosphere for this pulsar.

\subsection{PSR J1518+4904}

This pulsar, which has the longest period in our sample, is in a wide orbit
around a $\approx 1.0 M_{\sun}$ companion which is probably another
neutron star (\cite{nice96}). A post-cursor component has possibly
been identified (paper I) following the main pulse by 25$^{\circ}$.
The two features are connected with very low-level emission.  Moderate
levels of linear polarization and circular polarization that preserves
one sense are seen, while the PPA is well defined and has a shallow
slope. The spectral evolution of the profile components is {\em
negligible} between the 370-MHz discovery profiles and our measurements.
 
\subsection{PSR B1534+12}

The profile of this pulsar has been extensively studied by Arzoumanian
{\it et al.}~(1996) at 430 MHz and 1.4 GHz. The MP has a multiple
component shape with a total width of 7$^{\circ}$ at 10 \% of maximum,
including a surprisingly sharp component of $\approx 2.5 ^{\circ}$
width. With its moderate levels of linear polarization, and
sense-reversing circular polarization, this component is an
exceptionally narrow feature, unexpected in the context of a canonical
model. There is also low-level emission present throughout a
significant fraction of the period.  
The PPA curve is well defined,
and the orientation presented by Arzoumanian {\it et al.}~(1996) is
consistent with an orthogonal rotator. 
PSR B1534+12 is  a member of a NS-NS binary system and thus
one of the few cases where general relativity predicts a 
measurable precession. The precession rate predicted for the 
spin axis of B1534+12 amounts to  0.52$^\circ$yr$^{-1}$. 
Over the two years between our observations (April 1994) and those
presented by ~Arzoumanian {\it et al.}~(1996), such a small rate
would imply a change in the PPA curve slope of $\approx 1^{\circ}$,
assuming that the spin and orbital angular momentum axes of this
binary system are aligned. Though high signal-to-noise
observations are needed to detect precession, our
data do not present indications that  contradict the
predictions of relativity for this system.
We have used
the orientation of this pulsar to derive the emission altitude, and
estimate values of 30 and 47$R_{\ast}$ for the two solutions presented
by Arzoumanian {\it et al.}~(1996). There is {\em very little} change
in profile morphology with frequency with both the MP and the IP
fading very little relative to their conal outriders. The MP and IP do
not fade relative to each other with frequency. In normal pulsars, IPs
are rarely evident above 1.4 GHz.

\subsection{PSR B1620$-$26}

This pulsar is a member of a triple system (\cite{bac93}) located in
the globular cluster M4 (\cite{lyn88}). Its pulse profile is
symmetric, and perhaps the easiest to interpret in our sample since
the emission characteristics of the core component are clearly
evident.  The high degree of circular polarization associated with
significant linear polarization and a well defined PPA curve indicates
the existence of a core component between two conal outriders.
Comparing the 408-MHz (\cite{lyn88}), 790-MHz (\cite{bac95}), and
1300-MHz (\cite{fos91}) profiles with our 1.4-GHz measurements, there
is {\em no significant} frequency development in shape or width of the
profile.

Since B1620$-$26 was born in a globular cluster, it may not share the
same genesis as the field pulsars which dominate our sample. Despite
the expected modification of the field topology around the magnetic
axis by a prolonged accretion process (\cite{che93b}), we see no
evidence to support different observational characteristics for the
emission properties of this pulsar from the rest of the sample.

\subsection{PSR J1640+2224}

This MSP is a member of a binary system with a $\approx 0.3 M_{\sun}$
companion in a wide circular orbit (Foster {\it et al.}~1995). It
exhibits a wide, multi-component profile (paper I) with high linear
polarization and moderate circular polarization associated with the
profile centre. The PPA curve is strikingly flat.
In a sample of $\approx$ 300 polarimetrically studied normal pulsars
(Gould 1994),
five sources have been identified with PPAs flatter than 
$\left|\left(\frac{d\psi}{d\phi}\right)_{\rm max}\right|\le$0.5$\left(\frac{\rm deg}{\rm deg}\right)$.
These sources (e.g. PSR B0031-07, B0940+16, B1809-176, B1822-14 and
B2011+38)
have single component profiles which preserve their
simple shape
at higher frequencies, consistent with conal-single profile 
development. Conal-single profiles are associated with trajectories
of the line of sight that almost graze the emission cone. For these
five sources, fits to their PPA of a RVM predict a normalized
impact parameter between 0.94 and 0.98, which indeed signifies
very tangential trajectories. Despite the flat PPA curve,
PSR J1640+2224 does not share
a similar profile development with these sources. Instead, profile outriders
are evident at $\lambda$ 21cm, suggesting that the profile is consistent
with a core surrounded by a pair of conal outriders.
Triple profiles are associated with central trajectories
and hence much steeper slopes than the extremely flat one found here.
Thus, the flat PPA is difficult to understand within the same framework
which applies to normal pulsars. Mutlifrequency and high resolution
observations are essential in understanding this pulsar.
 
\subsection{PSR J1643$-$1224}

This source is among the most luminous MSPs known. It is a member of a
binary system, with a $\approx 0.14 M_{\odot}$ companion in a wide
circular orbit (\cite{lor95}).  The wide profile resembles the triple
core-dominated profile often seen amongst the normal pulsars. The
moderate degree of linear and significant circular polarization seen
here are properties often associated with single core component
profiles. This suggests that our line-of-sight cuts the polar cap very
close to the magnetic axis, although the PPA curve is rather flat for
such a geometry.  A very {\em slow} frequency development is noticed
for this profile.

\subsection{PSR J1713+0747}

This field MSP is in a wide circular orbit around a $\approx 0.3
M_{\sun}$ companion (\cite{fos93}).  The pulsar's large flux density
and sharp profile make it an excellent source for high-precision
timing experiments (\cite{cam94}). Kijak {\it et al.}~(1997) recently
reported the detection of this pulsar at 4.85 GHz.  At 1.4 GHz, the
profile can be separated into at least five components (paper I).
Sense-reversing circular polarization within the sharp central
component identifies this with the core component.  The leading part
of the profile, where additional components can be fitted, is only
weakly polarized, and the PPA curve is disturbed by 90$^\circ$
discontinuities.  Such discontinuities tend to correlate with sudden
drops of the linearly polarized intensity. These orthogonal emission
modes have long been held responsible for the depolarization of pulsar
radio emission. The PPA curve, when reconstructed for the orthogonal
mode, is rather flat.  At 430 MHz, the trailing part of the profile is
intrinsically broadened (\cite{fos93}). Between 1.4 and 4.85 GHz, no
significant profile development is observed, although the outermost
components are somewhat more prominent at 4.85 GHz than at lower frequencies.

\subsection{PSR J1730$-$2304}

This isolated MSP (\cite{lor95}) exhibits a multi-component profile at
1.4 GHz.  Changes in the profile shape, similar to the mode-changing
activity known from normal pulsars, have been identified (Kramer et
al. 1998b).  Different profile shapes are associated with changes
in the degree of linear polarization, ranging from completely
polarized to essentially unpolarized.  This peculiar behavior mainly
affects the observed linear polarization, while the circular
polarization appears stable. From the polarization profiles, we also
find sudden drops in the linear polarization which are associated with
90$^\circ$ discontinuities in the PPA curve. A small amount of
sense-reversing circular polarization seems to be associated with a
central core component, indicating that our line-of-sight cuts the
polar cap very close to the magnetic axis.  A post-cursor has been
detected (paper I), which follows the main pulse by $\approx 120
^{\circ}$. The two features are most likely not connected by low-level
emission, though future observations at lower frequencies might be
more informative.

\subsection{PSR J1744$-$1134}

This isolated MSP (\cite{bai96}) exhibits an interesting profile. High
signal-to-noise profiles presented in paper I reveal a pre-cursor
preceding the MP by $\approx 124 ^{\circ}$, as well as a trailing
component following the profile by $\approx 80^{\circ}$. Moderate
amounts of linear and circular polarization are seen. The PPA is well
defined and of steep slope compared with the slopes derived for the
majority of our sources. The PPA agrees well with the predictions of
a RVM (paper III).  {\em Very little} morphological change with
frequency is seen for this profile.

\subsection{PSR B1744$-$24A}
 
Located in the globular cluster Terzan 5, this is the second globular
cluster pulsar in our sample (cf.~PSR B1620$-$26). PSR B1744$-$24A is
an interesting source as its emission is eclipsed by a binary
companion (\cite{lmd+90}). The emission exhibits significant amounts
of linear polarization, whose degree is a function of orbital phase
and is varying between 60$\%$ and essentially zero. The associated
circular polarization remains constant (see ~Xilouris {\it et al.}
1998b {\em in prep.} for more details). The PPA curve is well defined with a
shallow slope.
  
\subsection{PSR B1855+09}

The detection of a Shapiro delay in the timing data of this relativistic
binary system (\cite{ryb91}) indicates its orbital plane is roughly
parallel to the line-of-sight, implying that PSR B1855+09 is most
likely an orthogonal rotator (\cite{tho90}). The separation 
$\Delta \phi_{\rm MP-IP}$ is $192^\circ$. However, the fit of a RVM to
the PPA curve (\cite{seg86}) does not provide conclusive evidence as
to whether the IP is produced at the same or the opposite magnetic
pole to the MP.  Comparing the $\Delta \phi_{\rm MP-IP}$ at 430
MHz (\cite{tho90}), 3 GHz (\cite{fos91}) and 4.85 GHz (\cite{kij97}),
shows that there is no significant change with frequency, suggesting
that the MP and the IP originate from two separate magnetic poles
(cf.~\cite{han86}).

The sense-reversing circular polarization detected under the central
part of the MP indicates the existence of a core component. A similar
effect also occurs for the IP, implying that both the MP and the IP
are created by a cut of the line-of-sight close to two different
magnetic poles, further supporting an orthogonal geometry.  We note
that the polarization properties of the interpulse derived from our
data are very similar to those presented by Segelstein {\it et al.}~(1986).  
The linear polarization does not drop to zero at
the pulse longitude where the PPA curve exhibits discontinuities, in
good agreement with the measurements of Segelstein {\it et al.}~(1986).
The discontinuities observed in our data are non-orthogonal,
which suggests that there is competition between modes of different
intensity. This could account for the slight change in the percentage
linear polarization observed among different sessions,
or shorter integrations of the same session.
The frequency development of the IP is
{\em unusual} since at higher frequencies it is stronger relative to
the MP (Kijak {\it et al.}~1997).  Extended low-level emission has
been reported only by ~Segelstein {\it et al.}~(1986). If confirmed,
this would be of interest since it cannot be reconciled with the
strong case presented here for an orthogonal rotator.

\subsection{PSR B1937+21}

Our observations of this pulsar are in agreement with previously
published polarimetry. However, due to hardware limitations, the time
resolution is poorer and we do not present them here and the following
discussion is based on the available literature.  The profile of PSR
B1937+21 consists of a MP, which has at least two components closely
spaced to each other, and an IP. The separation $\Delta \phi_{\rm
MP-IP} = 174^\circ$ remains constant for a wide frequency range, from
0.32 up to 2.38 GHz (see Ashworth {\it et al.} 1983; Stinebring \&
Cordes 1983; Cordes \& Stinebring 1984; Thorsett \& Stinebring 1990;
paper I).
The constant $\Delta \phi_{\rm MP-IP}$ could be perceived as an
indication that the emission originates from two different magnetic
poles. This view is further supported by the lack of low-level
emission for most of the period, at least with the current
observations (e.g.~Camilo 1995). The MP and the IP are both sharp
features, much narrower than the predictions of the canonical pulsar
model, while their width is strongly frequency dependent. The profile
development is normal and rather a weak function of frequency. The
amplitude ratio of the MP and IP, contrary to the behavior seen in
other MSPs with IPs, remains stable with frequency.  Following the
trend already recognized in other MSPs, the PPA curve of this star is
very flat across both the MP and the IP while signatures of orthogonal
moding are evident at the leading edge of the MP (e.g. ~Thorsett \&
~Stinebring 1990).  The linear polarization of the MP ranges from 64
\% at 430 MHz to 18 \% at 2.38 GHz (Thorsett \& ~Stinebring 1990).
Therefore, the emission properties of this pulsar are not
significantly different from those of a normal pulsar. Consequently,
its magnetosphere might not be deviating from a dipolar structure that
characterizes normal pulsars.  However, the sharpness of its
components remains a great puzzle.  Cordes \& Stinebring (1984)
suggest that the emission region of PSR B1937+21 must be very
compact. The lack of detectable non-dispersive time delays in pulse
arrival suggests that all emission from 0.3 to 1.4 GHz must arise from
the range of radii $\Delta r$=2 km.

Emission from a single pole, located very near the stellar surface,
has been suggested by Gil (1983) for this pulsar. In this model the
contribution of the quadrupolar field component to the total magnetic
field significantly alters the emission properties. Along similar
lines, complex field topologies in polar caps modified by accretion
during the spin-up or spin-down phase of the evolution of this star
have been held responsible for the sharp features and the flat PPA
curve observed (Chen \& Ruderman 1993b).  While the fashion in which
complex fields influence the emission remains unclear (e.g.~Krolik
1991), the emission properties of this pulsar are not so different
from normal pulsars.

\subsection{PSR B1953+29}

Although this pulsar was rather weak during our observations, we were
able to confirm the existence of the {\em pre-cursor} reported by
Thorsett \& Stinebring (1990).  This feature precedes the main pulse
by 120$^{\circ}$. Boriakoff {\it et al.}~(1986) found
no evidence for an interpulse at lower frequencies, while Stinebring
{\it et al.}~(1984) found no significant polarization.  Our
polarization measurements generally agree with those of Thorsett \&
~Stinebring (1990), although we observe some variability in the degree
of polarization. This could explain the non-detection of polarization
by ~Stinebring {\it et al.}~(1984).  Similarly, variability in
intensity could be responsible for the non-detection of the precursor
at 430 MHz.  The frequency development of this pulsar is certainly
{\em unusual} since the trailing part of the relatively simple profile
at 430 MHz evolves to a three-component profile at 1.4 GHz (paper I).

\subsection{PSR B1957+20}

The eclipsing binary pulsar B1957+20 exhibits the most puzzling
emission properties among all MSPs.  Despite our persistent efforts we
were unable to detect this source. However, for the sake of
completeness, a discussion of its properties is presented here based
on the available literature.  The profile of PSR B1957+20 consists of
a MP (nomenclature of ~Thorsett \& ~Stinebring 1990) which resembles a
core-single profile.  A rather broad and asymmetric IP is located
almost 180$^{\circ}$ from the MP. This component exhibits the most
unusual profile development among the known pulsars (Fruchter {\it et
al.}~1990).  Low-level emission connects these two components, while
two additional baseline-features have also been detected.
Both features precede the MP and the IP structure by about $40^{\circ}$
(Fruchter {\it et al.}~1990). While the MP resembles a
core-single profile, its frequency development is minimal (Thorsett
{\it et al.}~1989), in contrast to the predictions for this profile
class.  On the contrary, the frequency development of the IP does not
agree with the predictions of Rankin's classification
scheme. There is a substantial difference between the low and high
frequency profile of this component.  The baseline-features appear
stronger at low frequencies while their spectral development is
minimal as well.  The polarization properties of this pulsar remain
unknown and as suggested by Thorsett \& Stinebring (1990), they are
influenced by the environment of this eclipsing system. Theoretical
arguments strongly suggest that this pulsar is an orthogonal rotator,
making the origin of all the profile features an interesting puzzle.

Similar to B1937+21 the width of the MP of B1957+20 is much narrower
than expected. Therefore, the discussion presented for the profile of
B1937+21 might apply for the MP of B1957+20. However, this does not
apply to the IP. The properties of this component are totally
abnormal. The MP develops with frequency as expected in a traditional
polar cap environment.  On the contrary, the IP is far from this
description. Further observations are required to establish whether
the emission region of the IP might stem from a totally different
location than the MP. Its much wider profile suggests that the IP
might emanate from the vicinity of the light cylinder.  If this is the
case, then the abnormal profile properties should be associated with
emission from outer magnetospheric gaps.  The profile asymmetries of
the IP, as well as the extra baseline components could emerge as the
magnetic field lines at the vicinity of the light cylinder acquire an
azimuthal component, guiding the emission at wide angles from the IP.
The low degree of polarization detected at 430 MHz (Fruchter {\it et
al.}~1990) can not readily assist in verifying the origin of the MP
and IP of this source, however, such observations are needed to
clarify this issue.

\subsection{PSR J2019+2425}

This pulsar is a member of a binary system (\cite{nic93}) orbiting a
$\approx 0.37 M_{\sun}$ companion.  Due to severe scintillation, PSR
J2019+2425 is one of the most difficult sources to study. However, it
exhibits a perplexing profile and hence is included in our discussion
which is based on data presented by Nice (1992).  The MP is comprised
of two main components, A and B, following the nomenclature of Nice
(1992). A precursor (PC, component C) precedes the MP, while an almost
symmetrically spaced possible IP or postcursor (component D)
follows. Defining the profile center as the midpoint between component
A and B, the IP follows the MP by $\Delta \phi_{\rm MP-IP} \approx
123^\circ \pm 7^\circ$ at 430 MHz, a separation that marginally
narrows to $\approx 116 ^{\circ} \pm 7^\circ$ at 1.38 GHz. Some
evidence for the IP was also found in our 1.4 GHz data (paper I). The
PC precedes the MP by $\approx 104^{\circ}$ a separation that remains
constant over this frequency range.  It is worth noticing that both
the PC and the IP are $\approx$ 80\% weaker than the MP and are
located almost symmetrically around the profile midpoint. Whether
low-level emission connects each one of these components to the MP is
of great importance in interpreting this profile as single-pole or
double-pole emission. Further data with higher signal-to-noise are 
required to settle this issue.
  
The relative strength and separation of the PC and IP appear to be
frequency independent. At the same time, the MP shows a significant
change in shape between 430 MHz and 1.4 GHz, with its trailing part
(component A) becoming progressively weaker. Core components tend to
weaken with frequency, allowing for a pair of outriders around them to
emerge.  However, the MP of this profile exhibits only one conal
outrider (component B) making doubtful the association of component A
with a core component hence the profile development {\em abnormal}.
The MP of J2019+2425 could be a so--called partial cone, where only
one of the two conal outriders is present. Partial cones are a
relatively rare phenomenon, to date seen only amongst normal pulsars
(Lyne \& Manchester 1988).  However, the presence of two additional
components which are symmetrically spaced around the MP makes this
interpretation difficult, at least if the dominant field is a dipole.
Polarimetric data of high quality are required to shed light into the
interpretation of this profile.

\subsection{PSR J2051$-$0827}

This pulsar is a member of a binary system with an extremely low mass
companion ($\approx 0.03 M_{\sun}$) in a tight circular orbit which
eclipses the pulsar at radio frequencies below $\approx$ 1 GHz
(\cite{sta96}). It exhibits a simple profile that resembles those of
core single pulsars, with a moderate degree of linear and circular
polarization that preserves one sense throughout the profile.  The
low-level linear polarization seen here could be associated with the
presence of more components in this seemingly simple profile
(cf.~paper I).

\subsection{PSR J2145$-$0750}

This MSP (\cite{bai94}) is orbited by a $\approx 0.51 M_{\sun}$
white-dwarf companion in a wide circular orbit, which has been
identified optically at the timing position of the pulsar
(\cite{bel95}).  This pulsar exhibits a complex pulse morphology and
complicated polarization properties.  The profile consists of a main
pulse which closely resembles that of PSR B1237+25. However, it also
exhibits a precursor.  The precursor of the Crab pulsar is separated
from its MP by some 21$^{\circ}$ while a bridge of low-level emission
appears to connect the two features. The post-cursor of another normal
pulsar, PSR B0823+26, is separated from its MP by about 30$^{\circ}$,
again with low-level emission connecting the two features.  In
contrast, the precursor of PSR J2145$-$0750 is located relatively far
away ($\approx 96^{\circ}$) from the centre of the MP and it is an
unusual feature to interpret. At 1.4 GHz, we see no low-level emission
connecting the MP to the precursor.  We note that the precursor is not
always detectable in our shorter sub-integrations, as far as can be
inferred from the given signal-to-noise ratio.  At very low
frequencies (102 MHz), the precursor is not detected at all (Kuzmin \&
Losovsky 1996), while the separation of the precursor peak to the
leading edge of the MP remains unchanged between 430 MHz
(\cite{lor94}), 800 MHz (~Backer 1995), and 1.4 and 1.7 GHz (this
paper and paper I), suggesting emission from a single pole.  Under the
sharp leading component, the strange signature of both the linear and
circular polarization does not suggest any firm profile
classification.  The PPA curve deviates significantly from an RVM
curve, and is rather flat. Moreover, there is no indication in the
linear polarization that orthogonal mode-changing activity is present
which could alter the shape of the PPA curve. Therefore, the PPA curve
of this pulsar is hard to reconcile with a RVM curve, even if
disturbed by mode-changing activity.  The existence of a precursor at
an unexpected location, together with the severely perturbed PPA
curve, gives strong evidence for distortion of the magnetic field in
the emission regions of this source.

\subsection{PSR J2229+2643}

This pulsar is a member of a binary system in a wide circular orbit
with a $\approx 0.26 M_{\sun}$ companion. It exhibits the weakest
inferred surface magnetic dipole field strength in any known pulsar.
The 430-MHz profile (\cite{cam96c}) has {\em not evolved} by 1.4 GHz,
and profile narrowing with frequency is not seen.  The PPA is shallow,
and high linear and moderate circular polarization is evident.

\subsection{PSR J2317+1439}
  
PSR J2317+1439 is orbiting a $\approx 0.21 M_{\sun}$ companion in a
wide circular orbit (\cite{cam96c}).  The multiple-component profile
structure seen at 430 MHz indicates that the bright feature at the
central part of the profile is the core component. Indeed, at 1.4 GHz
this component has faded and become equal in intensity with the
leading component, {\em as expected} for a multiple-component profile.
There is significant circular polarization within the central part of
the profile, while the linear polarization remains at moderate levels.
The PPA curve exhibits a 90$^\circ$ discontinuity associated with a
drop of linearly polarized intensity.  There is no significant
narrowing of the profile width between 1.4 and 1.7 GHz.

\subsection{PSR J2322+2057}

PSR J2322+2057 is an isolated pulsar. Its relatively narrow, simple
components appear moderately polarized.  The $\Delta \phi_{\rm MP-IP}
= 234^\circ$ separation, remains unchanged with frequency to the
extent that it can be traced from the 430-MHz profile
(\cite{nic93}). As with PSR B1855+09 (Kijak {\it et al.}~1997), PSR
J2322+2057 shows a decreasing MP-IP intensity ratio with increasing
frequency {\em in contrast} to the general trend seen in the MP-IP
development for normal pulsars. The components are simple, but much
narrower than expected if a simple dipolar field structure is assumed
to prevail in the radio-emission region.

\section{Contrasting MSP and normal pulsar polarization properties}

It is evident from the discussion of Section 3, that the frequency
development of the profiles and the associated polarization
characteristics do not point towards a clear classification for most
of the MSPs studied.  Instead, we have identified pulsars for which
spectral profile development can be described as {\em normal} ({\bf
N}), {\em minimal} ({\bf M}) or {\em abnormal} ({\bf A}).  This
profile evolution is assigned to each source in Table 2 (col. 9), which
also lists the physical parameters that may regulate the radio
emission of each source.  A {\em minimal} development characterizes
profiles which exhibit virtually no change between frequencies of 400
MHz and 1.4 GHz. We characterize profiles as {\em normal} those which
develop according to the scheme of Rankin (1983a;b).  Finally,
{\em abnormal} development characterizes profiles which are
inconsistent with the predictions of Rankin's classification scheme.
The failure of the classification scheme to account for the emission
properties seen in MSPs could be attributed to either a change in the
emission mechanism itself for these much shorter rotational periods,
or to modifications in the local environment, namely the
magnetic-field structure in the emission region.

\placetable{tbl-1} 

A quantitative comparison is presented in this section between the
emission properties of millisecond and normal pulsars.  The
polarization parameters for the pulsars discussed above are summarized
in Table \ref{tbl-1}.  The mean linear (col.~2) and circular (cols.~3
and 4) polarization represent the ratio of the area under the $L$ or
$|V|$ (and $V$) curve respectively to the area under the $I$ curve
shown for each pulsar in Figs.~1-6.  The maximum fitted slope of the
PPA curve $\left(\frac{d\psi}{d\phi}\right)_{\rm max}$, is given in
col.~5, whenever its determination was possible.  The values listed in
Table \ref{tbl-1} are used to make a quantitative comparison between
the distribution of 1.4 GHz polarization parameters for normal and
MSPs.  In Fig.~7a, we present the histogram of the 1.4 GHz fractional
linear polarization for 281 normal pulsars from the sample of Gould
(1994) and for our measurements of 24 MSPs. This sample includes
the polarization of two MSPs with IPs.  It is evident that the
fractional linear polarization is higher on average at 1.4 GHz for
MSPs (Fig.~7a, upper panel) than for normal objects (Fig.~7a, lower
panel).  A similar effect is observed in the absolute value of the
circular polarization (Fig.~7b) with MSPs having statistically
significant higher polarizations than normal pulsars.  The histogram
of the values of circular polarization is presented in Fig.~7c where
the sense of circularity has been determined in our sample by
referencing our measurements to PSR B1929+10 whenever this was
possible. The deviation of the peak of this histogram from zero is at
2 $\sigma$ of its mean value.  This could be attributed to a systematic
instrumental effect but also suggests that the distribution presented
is incomplete.

The difference in the degree of polarization is further established by
the results of a Kolmogorov-Smirnov test that indicates a very low
probability (0.0004$\%$) that the two populations are similarly
polarized either in the linear or the circular component.  Our sample
includes half of the known MSPs with periods shorter than 30 ms, and
the results presented in Figs.~7a-c might change as more MSPs are
studied polarimetrically.  However, from this data set, evidence is
presented that MSP emission is more polarized than that of normal
pulsars at 1.4 GHz.

\begin{figure}
\plotfiddle{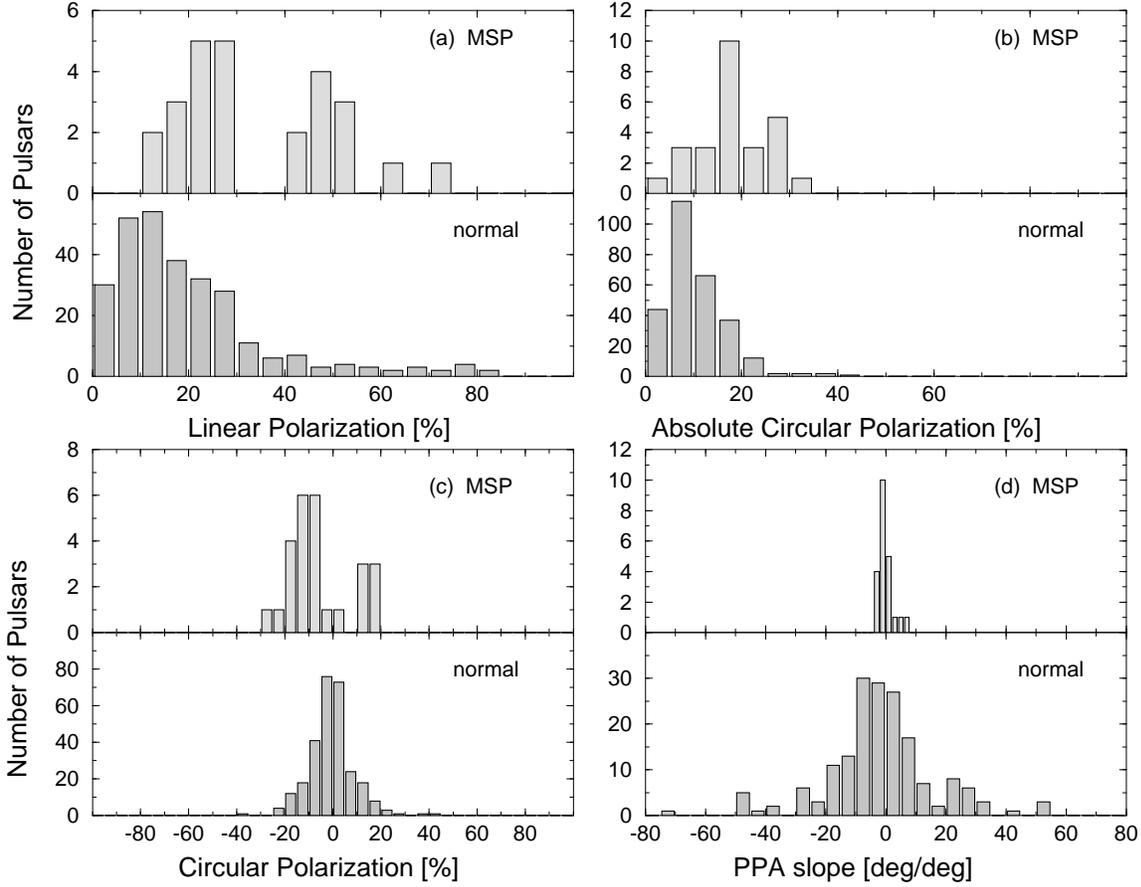}{10cm}{270}{60}{60}{-250}{350}
\figcaption{ \label{fig-7}
A statistical comparison of the emission properties between normal
pulsars and MSPs. The histogram of the fractional linear (a), absolute
circular (b) and circular (c) polarization is presented for 24 MSPs
(upper panel) and 281 normal pulsars (lower panel).
The polarization of the IPs of two MSPs in our sample is also
considered in these histograms.
 In (d) the
histogram of the steepest slope of the PPA curve for 20 MSPs in
our sample and two from the literature is
compared with the histogram for 178 normal pulsars.
}
\end{figure}

\placefigure{fig-7} 

Another result comes from comparing the distribution of the slopes of
the PPA curves for MSPs with that for normal pulsars in the Lyne \&
Manchester (1988) sample (Fig.~7d).  Even though some of our
observations (PSRs J1643$-$1224, B1744$-$24A and B1953+29) have been
influenced by dispersion smearing, we note that MSPs in general have
much shallower slopes than normal pulsars.  A Kolmogorov-Smirnov test
shows that the two distributions differ significantly.  A similar
comparison was also made with groups of normal pulsars which were
sorted according to their profile class. Our result is independent of
profile class.  It is evident from the polarization profiles
(Figs.~1-6) that the PPA curves of most MSPs, (a) cannot be easily
described by a RVM curve, (b) exhibit rather small excursions, and (c)
possess shallow slopes.

The polarization state of pulsar radio emission in a canonical pulsar
model is regulated by one or both of the following : (i) propagation
effects, and (ii) a geometrical overlapping of the elementary beams of
emission.  Deviations from the canonical model in MSP magnetospheres
might result in additional effects, modifying in some fashion the
polarization state (e.g.~Chen \& Ruderman 1993b).
 
One class of models of radio pulsar magnetospheres predicts an
outflowing electron-positron plasma through which radio-waves
propagate.  Propagation effects through this plasma should have
several observational consequences, as demonstrated by Cheng \&
Ruderman (1980) and extensively studied by Shitov (1985) and Barnard
(1986).  The propagation modes maintain a nearly fixed phase
relationship, and thus a constant polarization state at a certain {\em
limiting} radius. Barnard (1986) showed that under certain physical
conditions (depending on the rotational period $P$ and the magnetic
field strength $B$), this limiting radius approaches the light
cylinder radius ($R_{\rm LC}$) so that the field lines become quite
swept-back, exhibiting an increasingly azimuthal component. In this
case, the familiar S-shaped PPA curve will be modified (Barnard 1986,
Fig.~1).  The PPA will typically undergo a smaller swing and its curve
will exhibit a shallower slope due to the uniformity of the magnetic field
as a result of the azimuthal component contribution. The small
excursions in the PPA curves seen in the MSP profiles, together with
their shallower slopes make
these models a possible interpretation if indeed emission from close
to the light cylinder is an option.  Furthermore, emission at $R_{\rm
LC}$ would imply morphological asymmetries in the integrated profiles.
         
Alternatively, geometric arguments (Morris {\it et al.}~1981;~Xilouris
{\it et al.}~1995) relate the polarization state of the radio emission
to the size of pulsar magnetospheres, assuming processes such as
curvature radiation to be responsible for pulsar emission.
Significant fractional polarization evident in short-period pulsars is
consistent with the canonical pulsar model, in which relativistic
charges flow along dipolar field lines and emit curvature radiation
beamed tangentially to the field lines into narrow cones of opening
angle 1/$\gamma$, where $\gamma$ is the relativistic Lorentz
factor. The dipolar field lines, which control the motion of the
charged particles, diverge more for short-period pulsars at a given
radius than for long-period pulsars.  
Assuming that the elementary beams of emission are period independent,
they would then overlap less. This could therefore lead to a high
degree of linear polarization.  In contrast, converging beams would
lead to depolarization of the instantaneous radiation (see also Gil \&
Snakowski 1990).  While this geometric interpretation could account
for the somewhat higher fractional polarization seen in MSPs, it would
require a special geometry for the magnetic field in order to account
for disturbed PPA curves exhibiting shallow excursions and flat slopes.

Special field geometries could, for instance, emerge from modified
polar cap regions, such as those described by Chen \& Ruderman
(1993b). In the case of PSR B1937+21 they showed that the distorted
magnetic field would heavily influence the fractional linear
polarization and at the same time flatten out the PPA curve.  While a
firm identification of the core component in most of our MSP profiles
is rare, we point out that for these few cases the profiles preserve a
symmetry in their morphology.  In the presence of special field
geometries, surface emission models would result in complicated and
rather asymmetric profiles.  Such asymmetries are not evident in the
profile of PSR 1937+21 for instance.

The processes that regulate the degree of polarization, either
geometrical in origin or the result of propagation effects at the
light cylinder, do depend on the rotational period.  However, the
magnetospheric phenomena known to characterize the emission of normal
pulsars, such as orthogonal polarization modes, sense-reversing
circular polarization, or even intensity and polarization
mode-changing and intensity fluctuations, are clearly evident in the
MSP polarimetric profiles.  This strongly suggests that the basic
properties of the emission mechanism have not been influenced by the
change in rotational period.  Some unusual properties which we have
identified in the previous section could therefore be attributed to
environmental changes in the emission regions.  Such changes could be,
(a) special geometries due to mass accretion (e.g.~Chen \& Ruderman
1993b) or higher order magnetic field multipoles near the stellar
surface (e.g.~Manchester \& Johnston 1995), or (b) migration of the
location of the emission regions closer to the $R_{\rm LC}$
(e.g.~Barnard 1986).  A study of the frequency dependence of the
polarization properties of MSPs is eagerly awaited to clarify these
issues.

\section{The impact of the magnetic field on millisecond pulsar emission.}

So far we have shown that millisecond and normal pulsars most likely
share the same emission process, however, some MSPs exhibit unusual
properties which are not easily understood in the frame of a canonical
pulsar model. Certain physical parameters have been considered by
various magnetospheric models as factors that condition the
emission. Some of these parameters are given in Table 2 for the
pulsars in our sample.  In this section we investigate their role in
the regulation of the emission process.  At the same time we search
for common properties among the diversity exhibited by MSPs,
establishing in this way groups that may form the basis of future MSP
classification schemes.

\placetable{tbl-2} 

The distribution of $P$ (Table 2, col. 2) of MSPs exhibiting abnormal
profile development is relatively narrow, having an average value of
4.5 ms. In contrast, MSPs which exhibit normal characteristics
populate a much broader range as can also be seen in Fig.~8c.  The
dipole component of the surface magnetic field ($B_{\rm 0}$) can be
estimated using the $P$ and $\dot P$ of each pulsar.  The values of
$\dot P$, corrected for kinematic effects (cf.~Camilo {\it et
al.}~1994b) are used to estimate a lower limit ($\alpha =90^\circ$)
of the dipole component given in Table \ref{tbl-2} (col. 3).  We note
that pulsars exhibiting abnormal profile behavior are associated with
low values of $B_{\rm 0}$.  The respective distribution of MSPs with
normal characteristics is quite broad, while on average they tend
to exhibit higher fields.

A threshold parameter for the pulsar phenomenon is the potential drop
($\Delta V$) across the polar cap (e.g.~\cite{stu71}), given by
$\Delta V \approx B_{\rm 12} /P^{2}$ (Table 2, col.~4). This parameter
also constraints the size of sparks generated above a homogeneously
illuminated polar cap (Ruderman \& Sutherland 1975;~Cheng \& Ruderman
1980). A great scatter is evident among the values of $\Delta V$ for
the three classes of MSPs that we have identified (Fig.~8a).  However,
we note that most of them, particularly those with abnormal profile development
(marked with filled circles on Fig.~8a) exhibit relatively high values of the
accelerating potential.  High values of $\Delta V$ are representative
of normal pulsars classified as core-single (\cite{R93b}), where
$\Delta V$ is 2.5 on the average. Core-single stars are typically
young normal pulsars with narrow beams.  If indeed $\Delta V$
regulates in some fashion the radio emission then the abnormal
properties of MSPs might be encountered among the very young normal pulsars.

\cite{bes88} proposed the parameter $Q \approx 2P^{11/10}\dot
P^{-4/10}_{15} $, to distinguish between the propagation modes in
pulsar magnetospheres.  Values of $Q<$1 represent extraordinary
propagation modes while $Q>$1 are associated with normal propagation
modes of radiation.  According to the classification of some 150
pulsars (\cite{R93b}), values of the propagation parameter $1/Q >$ 1
are associated with the younger pulsars among the normal-pulsar
population.  Similar values seem to be representative of the MSPs in
our sample (see col. 5 of Table \ref{tbl-2}), suggesting common
properties of the polarization between the two groups.  Indeed,
core-single profiles are associated with moderate degree of linear and
significant circular polarization.  If the conditions in MSP
magnetospheres favour the presence of extraordinary modes, then the
high degree of circular polarization often identified in MSP emission,
might be the net result of the propagation of such modes.

The MSP magnetospheres are radially compact as can be seen from the
values of the light cylinder radii ($R_{\rm LC}$), given in col.~6 of
Table 2.  This suggests that any frequency stratification in their
magnetospheres (radius-to-frequency mapping; e.g.~Cordes 1978) should
be hardly evident in the pulse widths if we assume that the magnetic
field has preserved its dipole structure within the emission region.
MSPs exhibiting abnormal profile development in our sample tend to
exhibit the most compact magnetospheres with an average radius of
$R_{\rm LC}\approx$200 km.  They are also associated with wide polar
caps (see Table 2, col.~7).  Possibly, some of the abnormal effects
evident in MSP profiles might be related to higher order multipoles
acting in very radially compact but azimuthally extended
magnetospheres.  In contrast, sources in our sample that exhibit
minimal or normal profile development possess somewhat larger
magnetospheres though still compact if compared with normal pulsars.
The frequency dependence of the pulse widths of these MSPs suggests a
very minimal radius-to-frequency mapping in their magnetospheres, if
at all.  This implies that the presence of higher order multipoles is
rather limited. Indeed, a radius-to-frequency mapping acting in the
presence of higher order fields would result in more prominent profile
changes than those observed. However, it is also possible that
radius-to-frequency mapping scales in some fashion with $P$.
Multifrequency studies are needed to explore the period dependence of
radius-to-frequency mapping in MSPs.

Modifications are expected in MSP polar caps and hence the magnetic
field topology as a result of the accretion during the mass transfer
stage of their evolution (e.g.~Ruderman 1991;~Chen \& Ruderman 1993b).
Most likely this modification is regulated by the total mass
transferred and the duration of this process, while the initial
rotational period at the beginning of mass transfer might also act as
a tuning factor.  We note that the companion mass of MSPs with
abnormal properties in our sample is less than $0.4 M_{\sun}$
(col.~8). This associates them primarily with low mass companions,
mostly Helium white dwarfs.  In contrast, MSPs classified as normal or
minimal, are orbited both by Helium and Carbon-Oxygen companions.
High-mass binary pulsars are expected to have accreted a smaller
amount of mass than low-mass systems.  The preference of MSPs with
abnormal emission properties to be in low-mass systems suggests that
these stars might be associated with substantial mass transfer and
very short final rotational periods.  In paper I, a weak
anti-correlation was found between profile development and companion
mass as well as rotational period. Here we see that MSPs exhibiting
abnormal profile development are associated with substantial mass
transfer and a narrow range of periods.

We have computed the magnetic field at the light cylinder ($B_{\rm
LC}$) for all known MSPs, assuming a purely dipolar
magnetosphere. This is shown in Fig.~8b, as a function of period.
In Fig.~8b as well as in Figs.~8c and ~8d, 
MSPs with abnormal profile properties are identified by filled symbols.
Pulsars with IPs (PSRs J1012+5307, B1534+12, J1730$-$2304,
J1744$-$1134, B1821$-$24, B1855+09, B1937+21, B1957+20, J2019+2425, J2317+1439 and                                        
J2322+2057) as well as pulsars with components on their baselines
(pre- and post-cursors, i.e.~PSRs J1518+4904, J1730$-$2304, J1744$-$1134,
B1953+29, B1957+20, J2019+2425, J2145$-$0750 and J2317+1439) are
marked with triangles pointing up.  
Pulsars with extended emission
over their period or low-level emission at a wide range of longitudes
are shown as triangles pointing down (PSRs J0218+4232, J0437$-$4715
J0751+1807, J1012+5307, B1534+12, B1855+09, B1957+20, J2019+2425 and
J2317+1439). For comparison we also present the Crab (marked with a diamond)
and the Vela pulsar (marked with a square).
We note that very few sources  maintain a relatively high magnetic field
strength ($10^6$ G) at their light cylinder (Fig.~8b) which is similar to that
of the Crab pulsar. 
Among these sources are PSRs J0034$-$0534, J0218+4232, B1937+21 and B1957+20.
The IP of B1957+20 exhibits the most abnormal profile development seen
among MSPs while its MP as well as the components of PSR B1937+21 are
unusually narrow features.  Like the Crab pulsar,
B1937+21 exhibits unusual single pulses (\cite{cog96}).  
Although  the majority of MSPs maintains low magnetic fields near their
light cylinders,
many MSPs possess magnetic fields as found for the Vela pulsar
(i.e.~$10^4$ -- $10^5$ G) and
simultaneously exhibit an abnormal profile development.
In addition, we note that pulsars with IPs tend to form a boundary line, 
implying a period dependence of the existence of IPs on $\dot P$. 
Pulsars with extended emission also cluster along this boundary line.
Some magnetospheric models imply that the magnetic field at the
emission region of normal pulsars lies in the range
10$^6$ - 10$^7$ G (e.g.~Ruderman \& Sutherland 1975).
 This condition is satisfied at the vicinity of the light
cylinder by very few MSPs.  MSPs with abnormal profile
development as well as those with IPs or components of emission on
their baselines, satisfy this condition at a somewhat lower emission
height which however corresponds still to a significant fraction of their
light cylinder. Indeed, if a
field of $10^6$ G is essential for the creation of radio emission,
some MSPs could be potential outer gap emitters. However, MSPs with
unusually narrow components (e.g.~B1937+21), presumably due to some
current re-distribution, might meet this condition elsewhere in their
magnetospheres. Finally, there might be hybrid cases (e.g.~B1957+20)
where emission from outer gaps as well as traditional polar cap
emission might co-exist.

\begin{figure}
\plotfiddle{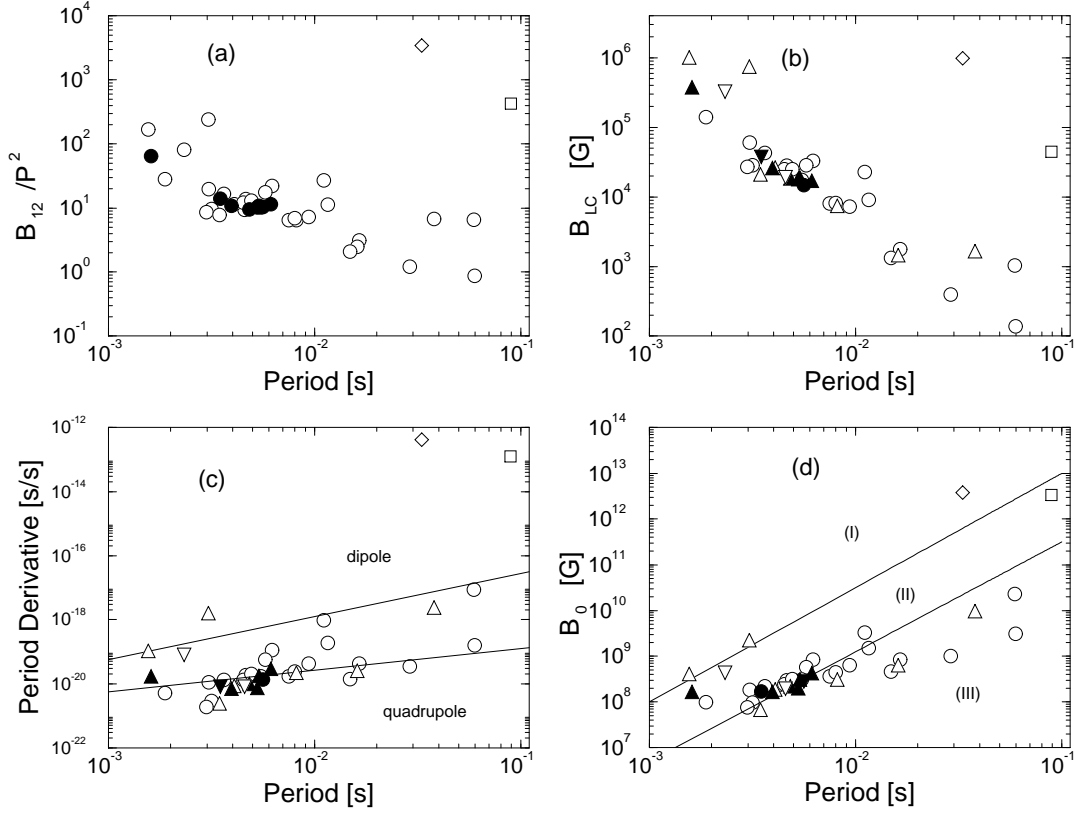}{10cm}{270}{60}{60}{-250}{350}
\bigskip

\figcaption{\label{fig-8}
Physical parameters of MSPs versus pulse period.
(a) accelerating potential, (b) dipolar magnetic field at the light
cylinder, (c) period derivative and (d) dipolar component of the
surface magnetic field.  Filled symbols
represent pulsars with abnormal profile development. 
In (b) -- (d) MSPs with IPs and pre or post cursors 
are marked with triangles pointing up, while sources with extended
emission are shown as triangles pointing down.
In (c) dipole and quadrupole spin-up lines are superimposed, while in
(d) the lines represent maximum locations of $\gamma$-ray emission
(see text for details).  
For comparison,
the Crab (diamond) and the Vela (square) pulsars are marked.}
\end{figure}

The distribution of MSPs on the $P-\dot P$ diagram has been attributed
to spin-up during disk accretion, if these stars have dipolar
magnetospheres. In Fig.~\ref{fig-8}c, we present the $P$-$\dot P$
diagram for all known MSPs. Superimposed on this are the spin-up lines
for dipole and quadrupole magnetospheres as calculated by Arons
(1993). Fig.~8c shows that, at the present stage of their evolution,
some of the MSPs are located near the dipolar spin-up line. However,
there is a substantial number of sources which presumably do not share
the same age or evolution, that cluster around the quadrupole spin-up
line. This spin-up line is calculated assuming that a quadrupole
field dominates the magnetosphere.
We note that abnormal MSPs are clustering around this 
spin-up line. If indeed these stars are associated with complicated
fields dominating their magnetospheres, they might exhibit a
somewhat higher value of the braking index.

While polar-cap models (e.g.~Ruderman \& Sutherland 1975) require
strong fields to sustain pair production, another series of models,
which involve emission from outer-gaps (e.g.~Cheng {\it et al.}~1986)
require somewhat reduced field strength.  Indeed, outer-gap emission
is strongly dependent on the magnetic-field strength, as seen from
Fig.~\ref{fig-8}d where the surface dipole fields of all known MSPs
are plotted against $P$.  According to the model of Chen \&
Ruderman (1993a), this diagram is separated into one region where the
traditional polar-cap emission prevails (region III), and two regions
where emission from outer gaps is possible (regions I and II).  Region
II includes pulsars which share common emission properties with the
Vela pulsar, while the properties that
correspond to pulsars populating region I are similar to the Crab
pulsar.  The death-line separating regions II and
III corresponds to the maximum of $\gamma$-ray emission which,
according to some authors, can occur only from the outer
magnetospheric gaps.  However, we should note that these lines are
strongly dependent on the assumed magnetic field structure
(e.g.~Phinney \& Kulkarni 1994). We note that 
pulsars with IPs and extended emission, populate
the border line that separates traditional polar-cap from outer-gap
emission.  According to a model by Romani \& Yadigaroglu (1995), there
is a variety of combinations of $\alpha$ and $\beta$ that would allow
the detection of this type of emission.  The highest probability
exists when an orthogonal rotator ($\alpha=90^\circ$) is viewed.
Our observation that most of the sources that fall on the border-line
are pulsars with IPs or exhibit extended emission, is in good
agreement with the predictions of this model.  This suggests that a
significant percentage of MSPs which combine IPs with extended
baseline emission could be understood as outer-gap emitters.  However,
pulsars such as B1937+21 exhibit IPs, lacking any observationally
confirmed low-level emission.  In addition, a simple scaling of the
canonical pulsar model fails to describe the narrow beams of the MP
and IP of this source (paper I).  This pulsar is then better
understood if the prevailing emission still originates from a
traditional polar cap which is not bound by the last open field lines,
but is rather contracted in size, forced by some kind of
redistribution of currents. Such a modification could be the product
of the accretion process responsible for the rebirth of this source
(Chen \& Ruderman 1993b).  However, it could also have its origins on
a pure geometrical contraction of the emission region, necessary for
its accommodation within the minute light cylinder of this pulsar at
regions where conditions favour radio emission.  Conceivably, this
contraction could account for the systematically narrower emission
beams evident in other MSPs as has been demonstrated in paper I.

\section{Conclusions}

To date, systematic surveys have revealed more than 40 pulsars with
periods $\le$ 30 ms (\cite{cam97}).  The emission properties of a
sizable number of these sources have been monitored for a period of
three years at a frequency of $\lambda$21 cm with the Effelsberg 100-m
radio telescope.  Our conclusions are presented in a series of
papers. In paper I, we presented the spectra and luminosities of MSPs
and investigated the profile complexity and the beam shape.
Millisecond pulsars are slightly less luminous sources, while their
spectra are identical to those of normal pulsars.  We have shown that
the complexity in MSP profiles is comparable to that seen in normal
pulsars. Their beam size does not follow the scaling predicted by a
canonical pulsar model. Rather, a critical period exists below which
the beams of MSPs appear narrower than expected.

In this work we have shown that the frequency development of the
polarization profiles and their properties do not comply with the
predictions of the classification schemes developed for normal
pulsars.  Instead, we can identify three classes of profiles: (a)
those that evolve minimally (M); (b) those that evolve as predicted
(N); and (c) those that evolve contrary to any prediction (A).  Eight
pulsars in our sample exhibit abnormal profile development and are
associated with short periods, low surface magnetic fields and compact
magnetospheres which could possibly engulf strong higher multipole
moments. In contrast, pulsars that evolve minimally or even normally,
are associated with somewhat larger and possibly dipolar
magnetospheres.  A very slow radius-to-frequency mapping must persist
in their magnetospheres, and is responsible for the minimal profile
change with frequency.

We find evidence that MSP emission is more polarized than that of
normal pulsars.  Moreover, the PPA curves of most MSPs, (a) cannot be
easily described by a RVM curve, (b) exhibit rather small excursions,
and (c) possess flat slopes.  These properties make models that
support emission from locations that constitute a substantial fraction
of the light cylinder (e.g.~Barnard 1986) a possible interpretation.
However, special geometries resulting from the binary nature of the
sources (e.g.~Chen \& Ruderman 1993b), or higher multipole moments
(e.g.~Manchester \& Johnston 1995), could influence the polarization
characteristics in a similar manner, prohibiting a clear
interpretation of the data, at least from the current single-frequency
observations.

Pulsars like B1937+21 and B1957+20 have beams which are narrower than
a simple scaling down of normal pulsar magnetospheres.  This
discrepancy can be reconciled if the emission in some MSPs still
originates from a traditional polar cap, but is not bound by the last
open field lines.  A possible redistribution of currents, either as
the result of an evolutionary effect or due to some scaling imposed by
$P$ and $\dot P$, could force the emission beam to become azimuthally
narrow.  This scenario could account for the narrow pulses of B1937+21
and the MP of B1957+20. However, the MP of B1957+20 exhibits different
properties that are better understood if its emission originates from
the vicinity of the light cylinder.  The profile asymmetries as well
as the extra components of emission on the baseline can be understood
in this model, as the net effect of field lines swept-back in the
azimuthal direction.

Evidently, the strength and topology of the magnetic field play an
important role in regulating the emission in MSP magnetospheres.  Some
sources maintain a strong dipolar field (10$^6$ G) at locations which
constitute a significant fraction of their light cylinder.  Among
those, some pulsars with IPs and extended emission can be found.
Pulsars with abnormal profile development also seem to radiate from
such locations.

Although the fashion in which the presence of higher multipole moments
influences the profile shape and the polarization characteristics is
uncertain, we note that pulsars with IPs that show abnormal properties
cluster around the quadrupolar spin-up line in the $P-\dot P$
diagram. If indeed the magnetospheres of these stars are quadrupolar
this implies that they might exhibit a higher than the average braking
index.

The combination of $B_{\rm 0}$ and $P$ of some pulsars with IPs and
extended emission places these sources near the death-line that
separates outer-gap emission from traditional polar-cap emission.
According to certain models (e.g.~Romani \& Yadigaroglu 1995), the
highest probability of detecting emission from outer magnetospheric
gaps exists for an orthogonal rotator.  This prediction is in very
good agreement with our findings, suggesting that the low-level
emission and the unusual features identified in the baselines of some
MSPs could result from outer gaps.  However, not all pulsars with IPs
fit in this scheme since extended emission at low-levels is not
identified (e.g~B1937+21), nor are all pulsars with IPs abnormal in
their frequency evolution.  Depending on certain physical or
geometrical parameters and possibly below a critical rotational
period, the emission region could have migrated into locations where
more favourable conditions exist.

Millisecond pulsar magnetospheres are indeed compact and the
distinction between inner and outer magnetospheric regions might not
be as clear as in the case of normal pulsars. It could be that the
emission is regulated by a combination of processes where favourable
conditions decide on the prevailing process.  The work presented in
this paper represents first attempt to understand the emission of MSPs
based on a large sample of objects.  Multi-frequency polarimetric and
timing observations are essential in shedding light on the issue of
the location of the emission region and the structure of the magnetic
field in MSP magnetosphere.

\acknowledgments 
We are thankful to C. ~Salter, J. ~Eder and Z. ~Arzoumanian for their
 helpful comments. We are also grateful to the unknown referee for
 his helpful comments.
FC gratefully acknowledges the support of the European Commission
through a Marie Curie fellowship, under contract nr.~ERBFMBICT961700.
This work was in part supported by the European Commission under the
HCM Network Contract nr.~ERB CHRX CT960633, i.e.~the {\it European
Pulsar Network}.
Arecibo Observatory is operated by Cornell University under
cooperative agreement with the ~National ~Science ~Foundation.

\clearpage

\newpage

\begin{deluxetable}{lr@{$\; \pm \;$}lr@{$\; \pm \;$}lr@{$\; \pm \;$}lr@{$\;
\pm \;$}l}
\small
\tablenum{1}
\tablewidth{0pt}
\tablecaption{\label{tbl-1} Polarization parameters for 24 MSPs. 
The mean fractional polarization (linear, absolute circular and
circular) and the associated errors are presented in cols. 2, 3, and 4
respectively. The maximum slope of the PPA curve is presented in
column 5}
\tablehead{
\colhead{PSR}     & 
\multicolumn{2}{c}{$\frac {L}{I}$} &
\multicolumn{2}{c}{$\left | \frac {V}{I}\right |$}&
\multicolumn{2}{c}{$\frac {V}{I}$}&
\multicolumn{2}{c}{$\left(\frac{d\psi}{d\phi}\right)_{\rm max}$}\\
\colhead{}     & 
\multicolumn{2}{c}{(\%)} &
\multicolumn{2}{c}{(\%)} &
\multicolumn{2}{c}{(\%)}& 
\multicolumn{2}{c}{$\left(\frac{\rm deg}{\rm deg}\right)$} 
}
\startdata
J0613$-$0200&26.3&2.1 &16.2&1.1&$-$12.7&1.1   & \multicolumn{2}{c}{\nodata}\nl
J0621+1002 &14.7&1.5 & 7.6&1.3 &$-$6.7&1.3    & \multicolumn{2}{c}{\nodata}\nl 
J0751+1807 &28.5&1.5 &12.7 &1.0 &$-$9.3&1.0    &1.86 &0.19 \nl 
J1012+5307 &54.8&0.7 &18.4&0.6 &$-$11.3&1.0  &$-$2.70&0.09   \nl
J1022+1001 &52.8&0.4 &15.7&0.4 &$-$12.6&0.4   &6.12&0.20 \nl 
\tablevspace{5pt}
J1024$-$0719&47.8&1.7 &18.3&1.4 &$-$9.6&0.4    &0.34&0.03 \nl 
B1257+12   &51.7&2.9 &25.0&2.7  &$-$17.8&2.7   &$-$1.17&0.08   \nl 
J1518+4904 &21.7&1.7 &19.5&1.7  &16.5 &1.7   &$-$1.32&0.21 \nl
B1534+12   &17.9&2.7 &15.1&4.5  &$-$5.1&4.5   &$-$2.00&0.19  \nl
B1620$-$26  &20.8&1.6 &21.0&1.6 &13.5&1.6    &$-$1.52&0.10  \nl 
\tablevspace{5pt}
J1640+2224 &70.3&0.8 &27.1&0.4 &$-$21.9&0.4   &0.17&0.02   \nl
J1643$-$1224&22.8&1.0 &21.3&0.8 &$-$10.8&0.7   &$-$1.10&0.06 \nl
J1713+0747 &29.7&0.2 &15.5&0.2 &15.3&0.2    &$-$1.70&0.04   \nl
J1730$-$2304&60.9&1.0 & 4.7&0.5 &10.4&0.4    &2.23&0.09   \nl
B1744$-$1134&26.8&1.7 & 9.6&1.3 &$-$9.1&1.3     &$-$3.70&0.12    \nl
B1744$-$24A &46.1&2.0 &12.3&1.3 &$-$8.2&1.5   &$-$0.40&0.03    \nl
\tablevspace{5pt}
B1855+09MP  &19.5&0.5 &13.6&0.5&12.3&0.5    &$-$1.67&0.07   \nl
B1855+09IP  &15.8&1.7 &22.2&1.9&18.0&1.9    &$-$1.14&0.48    \nl
B1937+21MP &24.6&0.1  & 6.5&0.1&0.1 &0.1    & \multicolumn{2}{c}{\nodata}\nl
B1937+21IP &46.9&0.1  &11.3&0.1&$-$0.5&0.1    &\multicolumn{2}{c}{\nodata}\nl 
B1953+29   &25.7&2.8  &27.7&1.6&$-$5.3&2.2    &1.38&0.19    \nl
\tablevspace{5pt}
J2051$-$0827&11.9&2.1  &25.6&2.1&$-$13.7&2.2   &$-$1.36&0.49 \nl
J2145$-$0750&45.1&0.3  &17.7&0.2&$-$17.6&0.2   &$-$0.65&0.04  \nl
J2229+2643 &42.3&2.6  &17.6&1.0&$-$15.3&1.5  &$-$0.55&0.08   \nl
J2317+1439 &24.1&0.8  &25.7&0.4&$-$25.7&0.9  &1.65&0.06   \nl
J2322+2057MP&27.6&4.6 &16.7&3.8&$-$12.1&3.5   &\multicolumn{2}{c}{\nodata}\nl
J2322+2057IP&41.6&4.6 &34.7&3.8&$-$15.1&3.5   &\multicolumn{2}{c}{\nodata}\nl
\enddata
\end{deluxetable}
\clearpage
\begin{deluxetable}{lrrrrrrcc}
\small
\tablenum{2}
\tablewidth{0pt}
\tablecaption{\label{tbl-2} Physical Parameters of  MSPs.}
\tablehead{
\colhead{PSR}     & \colhead{$P$} &
\colhead{$B$}& \colhead{$\left(\frac{B_{\rm 12}}{P^{2}}\right)$}&
\colhead{$\left(\frac{1}{Q}\right)$}& \colhead{$R_{\rm LC}$}&
\colhead{$R_{\rm cap}$} & \colhead{$m_{\rm 2}$}&
\colhead{Profile}\\
\colhead{}     & \colhead{(ms)} &
\colhead{(10$^{8}$ G)}& \colhead{}&
\colhead{}& \colhead{(km)}&
\colhead{(km)} & \colhead{(M$_\odot$)} &
\colhead{evolution}\\
}
\startdata
J0613$-$0200& 3.062&   1.86&  19.83&   3.03& 146.2& 2.71 & 0.15  & M/A? \nl
J0621+1002&  28.854&  10.17&   1.22&   0.41&1377.7& 0.88 & 0.54  &M \nl
J0751+1807&   3.479&   1.69&  13.95&   2.32& 166.1& 2.54 & 0.15  &A \nl
J1012+5307&   5.256&   2.80&  10.15&   1.87& 251.0& 2.07 & 0.13  &A \nl
J1022+1001&  16.453&   8.41&   3.11&   0.81& 785.6& 1.17 & 0.87  &N \nl
\tablevspace{5pt}
J1024$-$0719& 5.612&   3.25&  10.32&   1.91& 268.0& 2.00& \nodata&A \nl
B1257+12&     6.219&   8.49&  21.96&   3.53& 296.9& 1.90 & \nodata&M/N \nl
J1518+4904&  40.935&  \nodata&\nodata&\nodata&1954.5& 0.74 & 1.01  &M \nl
B1534+12&    37.904&  97.12&   6.76&   1.65&1809.8& 0.77 & 1.34  &N \nl
B1620$-$26&   11.076&  33.18&  27.04&   4.41& 528.9& 1.43 & \nodata &M \nl
\tablevspace{5pt}
J1640+2224&   3.163&   0.97&   9.69&   1.71& 151.0& 2.67 & 0.30  &M \nl
J1643$-$1224& 4.621&   2.96&  13.86&   2.37& 220.6& 2.21 & 0.14  &M \nl
J1713+0747&   4.570&   1.96&   9.36&   1.73& 218.2& 2.22 & 0.33  &N \nl
J1730$-$2304& 8.123&   4.25&   6.44&   1.36& 387.9& 1.66 & \nodata&M \nl
J1744$-$1134& 4.075&   1.89&  11.41&   2.00& 194.6& 2.35 & \nodata&M \nl
\tablevspace{5pt}
B1744$-$24A&  11.564&  15.00&  11.22&   2.19& 552.2& 1.39 & \nodata&M \nl
B1855+09  &    5.362&   3.13&  10.89&   1.98& 256.0& 2.05& 0.26  & A \nl
B1913+16&    59.030& 228.36&   6.55&   1.68&2818.5& 0.62 & 1.39  &N \nl
B1937+21&     1.558&   4.10& 168.70&  15.70&  74.4& 3.80 & \nodata&M \nl
B1957+20&     1.607&   1.66&  64.38&   7.29&  76.7& 3.74 & 0.03  &A \nl
\tablevspace{5pt}
B1953+29&     6.133&   4.30&  11.44&   2.09& 292.8& 1.92 & 0.21  &A \nl
J2019+2425&   3.935&   1.68&  10.86&   1.92& 187.9& 2.39 & 0.37  &A \nl
J2051$-$0827&  4.509&   2.45&  12.05&   2.11& 215.3& 2.23 & 0.03  &M \nl
J2145$-$0750& 16.052&   6.41&   2.49&   0.68& 766.4& 1.18 & 0.51  & M \nl
J2229+2643&   2.978&   0.76&   8.58&   1.55& 142.2& 2.75 & 0.15  &M \nl
\tablevspace{5pt}
J2317+1439&   3.445&   0.92&   7.79&   1.45& 164.5& 2.56 & 0.21  &N \nl
J2322+2057&   4.808&   2.19&   9.47&   1.75& 229.6& 2.16 & \nodata&A \nl
\enddata
\end{deluxetable}

\end{document}